\documentclass[fleqn,usenatbib]{mnras}

\usepackage{newtxtext,newtxmath}

\usepackage[T1]{fontenc}

\usepackage{ulem}

\DeclareRobustCommand{\VAN}[3]{#2}
\let\VANthebibliography\thebibliography
\def\thebibliography{\DeclareRobustCommand{\VAN}[3]{##3}\VANthebibliography}

\usepackage{graphicx}	
\usepackage{amsmath}	

\usepackage{multirow}
\usepackage{array}

\usepackage{booktabs,threeparttable}
\usepackage{comment}

\usepackage{xcolor}

\newcommand{\msun}{\mbox{M$_{\odot}$}}
\newcommand{\kms}{\mbox{$\rm{km}\,s^{-1}$}}
\newcommand{\gcm}{\,g\,$\mathrm{cm}^{-3}$}
\newcommand{\ergscmA}{erg\,s$^{-1}$\,cm$^{-2}$\,\AA$^{-1}$}

\newcommand{\tardis}{$\textsc{tardis}$}
\newcommand{\carsus}{$\textsc{carsus}$}
\newcommand{\paperII}{\mbox{Paper II}}

\newcommand{\I}{\,{\sc i}}
\newcommand{\II}{\,{\sc ii}}
\newcommand{\III}{\,{\sc iii}}
\newcommand{\IV}{\,{\sc iv}}
\newcommand{\SrII}{Sr\,{\sc ii}}
\newcommand{\CaII}{Ca\,{\sc ii}}
\newcommand{\MSrII}{$M_{\rm{Sr}\,\textsc{ii}}$}
\newcommand{\YII}{Y\,{\sc ii}}
\newcommand{\ZrII}{Zr\,{\sc ii}}
\newcommand{\CeII}{Ce\,{\sc ii}}
\newcommand{\NdII}{Nd\,{\sc ii}}
\newcommand{\NdIII}{Nd\,{\sc iii}}
\newcommand{\SmII}{Sm\,{\sc ii}}
\newcommand{\EuII}{Eu\,{\sc ii}}

\newcommand{\rpro}{\mbox{$r$-process}}
\newcommand{\Aval}{\mbox{$A$-values}}
\newcommand{\gfo}{AT2017gfo}
\newcommand{\xsh}{\mbox{X-shooter}}
\newcommand{\Xlanth}{$X_{\textsc{ln}}$}

\newcommand{\Ye}{$Y_{\rm e}$}
\newcommand{\AngI}{\Ye$- 0.44$a}
\newcommand{\AngXII}{\Ye$- 0.44$b}
\newcommand{\AngII}{\Ye$- 0.37$a}
\newcommand{\AngXI}{\Ye$- 0.37$b}
\newcommand{\AngIII}{\Ye$- 0.29$a}
\newcommand{\AngX}{\Ye$- 0.29$b}
\newcommand{\AngIV}{\Ye$- 0.21$a}
\newcommand{\AngIX}{\Ye$- 0.21$b}
\newcommand{\AngV}{\Ye$- 0.12$a}
\newcommand{\AngVIII}{\Ye$- 0.12$b}
\newcommand{\AngVI}{\Ye$- 0.05$a}
\newcommand{\AngVII}{\Ye$- 0.05$b}

\title[Modelling the photospheric spectra of \gfo]{Modelling the spectra of the kilonova \gfo\ -- I: The photospheric epochs}

\author[J. H. Gillanders et al.]{
J. H. Gillanders$^{1}$\thanks{E-mail: jgillanders01@qub.ac.uk},
S. J. Smartt$^{1}$,
S. A. Sim$^{1}$,
A. Bauswein$^{2,3}$, 
S. Goriely$^4$
\\
$^{1}$Astrophysics Research Centre, School of Mathematics and Physics, Queen’s University Belfast, BT7 1NN, UK\\
$^2$GSI Helmholtzzentrum f\"ur Schwerionenforschung, Planckstrasse 1, 64291 Darmstadt, Germany\\
$^3$Helmholtz Research Academy Hesse for FAIR (HFHF), GSI Helmholtz Center for Heavy Ion Research, Campus Darmstadt, Germany\\
$^4$Institut d'Astronomie et d'Astrophysique, CP-226, Universit\'{e} Libre de Bruxelles, 1050 Brussels, Belgium\\
}

\date{Accepted XXX. Received YYY; in original form ZZZ}

\pubyear{2021}

\begin{document}
\label{firstpage}
\pagerange{\pageref{firstpage}--\pageref{lastpage}}
\maketitle

\begin{abstract}
The kilonova (KN) associated with the binary neutron star (BNS) merger GW170817 is the only known electromagnetic counterpart to a gravitational wave source. Here we produce a sequence of radiative transfer models (using \textsc{tardis}) with updated atomic data, and compare them to accurately calibrated spectra. We use element compositions from nuclear network calculations based on a realistic hydrodynamical simulation of a BNS merger. We show that the blue spectrum at +1.4 days after merger requires a nucleosynthetic trajectory with a high electron fraction. Our best-fitting model is composed entirely of first $r$-process peak elements (Sr \& Zr) and the strong absorption feature is reproduced well by Sr\,\textsc{ii} absorption. At this epoch, we set an upper limit on the lanthanide mass fraction of $X_{\textsc{ln}} \lesssim 5 \times 10^{-3}$. In contrast, all subsequent spectra from $+2.4 - 6.4$ days require the presence of a modest amount of lanthanide material ($X_{\textsc{ln}} \simeq 0.05^{+0.05}_{-0.02}$), produced by a trajectory with $Y_{\rm e} = 0.29$. This produces lanthanide-induced line blanketing below 6000\,\AA, and sufficient light $r$-process elements to explain the persistent strong feature at $\sim 0.7 - 1.0$\,\micron\ (Sr\,\textsc{ii}). The composition gives good matches to the observed data, indicating that the strong blue flux deficit results in the near-infrared (NIR) excess. The disjoint in composition between the first epoch and all others indicates either ejecta stratification, or the presence of two distinct components of material. This further supports the `two-component' kilonova model, and constrains the element composition from nucleosynthetic trajectories. The major uncertainties lie in availability of atomic data and the ionisation state of the expanding material.
\end{abstract}


\section{Introduction} \label{sec:Introduction}
\par
Mergers of binary neutron star systems have long been mooted as an ideal location for the synthesis of rapid neutron capture (\rpro) elements \citep[][]{Metzger2017}. Many different theoretical simulations have shown that the extreme ejecta properties, coupled with the large neutron fraction in the expelled material is sufficient for the \rpro\ to be viable \citep[][]{Lattimer1974, Eichler1989, Freiburghaus1999, Rosswog1999, Goriely2011, Goriely2013, Goriely2015, Korobkin2012, Perego2014, Wanajo2014, Just2015}. However, spectrophotometric observations of real mergers are needed to confirm the predicted properties of these models.

\par
The first discovery of gravitational waves from a BNS merger was made by the LIGO--Virgo scientific collaboration in 2017 \citep[GW170817;][]{LigoVirgo2017}. This led to the discovery of the kilonova \gfo, the electromagnetic counterpart to the BNS merger \citep[see][]{LigoVirgo2017, Andreoni2017, Arcavi2017, Chornock2017, Coulter2017, Cowperthwaite2017, Drout2017, Evans2017, Kasliwal2017, Lipunov2017, Nicholl2017, Pian2017, Shappee2017, SoaresSantos2017, Smartt2017, Tanvir2017, Troja2017, Utsumi2017, Valenti2017}. There have been numerous attempts to interpret the evolution of the early spectra and photometry of \gfo, with many authors invoking two components.

\par
In a two-component model, two distinct components of ejecta material are produced. First, ejecta material is expelled dynamically, either during or immediately after the merger (on $\sim$~millisecond timescales). This is followed by a second component of ejecta material (on timescales of $\sim$~a second), that is partially blown away from a newly formed accretion disk surrounding the merged remnant \citep[see][or any of the subsequent references, for more on the two-component model]{Kasen2017}. The expected mass of dynamical ejecta from simulations of BNS mergers lies in the region \mbox{$10^{-4} < M_{\rm dyn} < 10^{-2}$\,\msun} \citep[e.g.][]{Ruffert1996, Rosswog1999, Goriely2011, Korobkin2012, Bauswein2013, Hotokezaka2013, Palenzuela2015, Foucart2016, Sekiguchi2016, Ciolfi2017, Radice2018, Ardevol2019, Nedora2019, Just2022b, Kullmann2022}, with typical ejecta speeds in the range \mbox{$0.2 < v < 0.3\,c$}. The disk wind can be more massive (\mbox{$10^{-2} < M_{\rm{dw}} < 10^{-1}$\,\msun}) and is slower moving \citep[\mbox{$v \sim 0.1\,c$}; e.g.][]{Fernandez2013, Metzger2014, Perego2014, Just2015, Wu2016, Siegel2017, Fujibayashi2018, Miller2019, Curtis2021, Just2022}. The two-component model has been used by numerous authors to explain the spectral energy distribution (SED) of the early spectra of \gfo\ \citep[see][]{Chornock2017, Cowperthwaite2017, Kasen2017, Coughlin2018}. \cite{Perego2017}, \cite{Villar2017} and \cite{Breschi2021} have gone further, and suggested that three components with different opacities may be required.

\par
BNS mergers generally provide very favorable conditions for \rpro\ element nucleosynthesis, and models can closely reproduce the solar \rpro\ abundance pattern \citep[e.g.][]{Freiburghaus1999, Goriely2011, Korobkin2012, Bauswein2013, Perego2014, Wanajo2014, Just2015, Radice2018, Ardevol2019, Just2022, Kullmann2022}. Main challenges include the modelling of the different mass ejection channels of a merger and, in particular, the exact distribution of the electron fraction (\Ye) in the different ejecta components, which strongly affects the range of \rpro\ elements being produced. Also, different nuclear physics models determining the reaction and decay rates of the \rpro\ lead to recognizable variations of the elemental abundances and associated decay heating \citep[e.g.][]{Goriely2013,Goriely2015b, Mendoza2015,Martin2016,Giuliani2018,Lemaitre2021,Zhu2021}.

\par
Radiative transfer spectral modelling works focussed on \gfo\ have broadly followed two paths. The first has been to try to directly identify features in the early spectra that belong to specific elements, and use these to infer information about the ejecta material. \cite{Smartt2017} suggest that the early spectral absorption can be attributed to Te\I\ and Cs\I, which are both second \rpro\ peak elements. Further analysis by \cite{Watson2019} suggests that the early spectra of \gfo\ can be explained by lighter \rpro\ material. Specifically, they attribute the same absorption feature in the early spectra to \SrII, a first \rpro\ peak element. Both studies point to the presence of \rpro\ material, although they disagree over which elements are dominating the early spectra. \cite{Gillanders2021} searched for signatures of platinum and gold in the spectra, using new atomic data from \cite{McCann2022}, but no definitive identification was forthcoming. \cite{Domoto2021} model the entire KN ejecta, and find that the \gfo\ spectra can be reproduced with a lanthanide-poor ejecta composition, with the early spectral absorption dominated by \SrII\ \citep[corroborating the finding of][]{Watson2019}. \cite{Perego2022} explore the production of light elements ($Z < 20$) and Sr in early-phase KN ejecta, and present some comparisons to \gfo. They show that He may contribute to absorption in the early spectra of \gfo, although the mass needed to produce a prominent feature exceeds the predicted synthesised mass of He by around an order of magnitude. The predicted synthesised mass of Sr is in agreement with the Sr mass inferred by \cite{Watson2019}.

\par
The second approach has been to generate theoretical atomic data and use those to model the broad spectral shapes. There are a number of different groups working on theoretical atomic data that can be used to investigate BNS mergers \citep[e.g.][]{Japan-Lithuania_opacity_database, NIST-LANL}. Models presented by \cite{Kasen2017} and \cite{Tanaka2018, Tanaka2020} show that the SEDs produced from low \Ye\ material can reproduce the rising NIR flux observed in the evolution of \gfo. They show that the presence of lanthanides in particular critically affect the shape of the spectrum, resulting in a significant shift of the emission to the NIR. Since these studies use atomic data that are generally not calibrated to laboratory observed transitions, there will be some systematic uncertainty in both the wavelength and strengths of transitions in their models. Line blending due to the high velocities and the multitude of transitions of these elements also add to the difficulty in clear identification of particular ions. While the \cite{Kasen2017} and \cite{Tanaka2018, Tanaka2020} models are valid for modelling the overall shape of the spectra and inferring the likely element mix, they are less effective in providing direct line identification. A further complication is understanding the physics of the radioactive heating and thermalisation mechanisms \citep{Barnes2021}. \cite{Fontes2020} tabulated wavelength-dependent opacities for all lanthanides and uranium, which have been used by several authors \citep{Even2020, Wollaeger2021, Korobkin2021} to generate SEDs to compare with \gfo, but this method also does not allow identification of specific transitions.

\par
To date, there is no published model that reproduces all the high quality, accurately calibrated, daily spectra of \gfo\ consistently with a physically plausible mix of \rpro\ elements. In this work, we use such an element mix from nucleosynthetic trajectories in BNS mergers, and calculate new radiative transfer models with all publicly available (mostly calibrated) atomic data for the elements beyond the iron group. Our spectral models reproduce the overall shapes of the observed spectra, and we also identify the atoms and ions that have the most prominent effects (absorption/emission) on the spectra. In Section~\ref{sec:The observed spectral sequence of AT2017gfo} we motivate our choice of spectral data to model. In Section~\ref{sec:Atomic data}, we present our compiled atomic data set, and in Section~\ref{sec:Composition profiles} we present the composition profiles from hydrodynamic nucleosynthesis calculations. In Section~\ref{sec:Spectral analysis method}, we describe the methodology for our models, and present the main results in Section~\ref{sec:Spectral analysis results}. We follow up with some discussion and our interpretation in Section~\ref{sec:Discussion and interpretation}. Finally, we conclude in Section~\ref{sec:Conclusions}.

\par
This paper focuses on the spectra which we consider to be still within the photospheric phase and therefore valid for \tardis\ modelling, as described in Section~\ref{sec:Spectral analysis method}. A second paper (Gillanders et al. 2022, in prep.) will analyse the spectra taken $\gtrsim 7$ days. We refer to this as \paperII\ throughout this manuscript.

\section{The observed spectral sequence of \gfo} \label{sec:The observed spectral sequence of AT2017gfo}
\par
Throughout this work, we present comparisons of models to the observed spectra of \gfo. We primarily use the set of ten \xsh\ spectra originally published by \cite{Pian2017} and \cite{Smartt2017}, and re-reduced and re-calibrated by the ENGRAVE collaboration \citep{ENGRAVE2020}. These spectra have been flux-calibrated to a compiled set of photometric measurements taken from the published values of \cite{Andreoni2017, Arcavi2017, Chornock2017, Cowperthwaite2017, Drout2017, Evans2017, Kasliwal2017, Pian2017, Smartt2017, Tanvir2017, Troja2017, Utsumi2017, Valenti2017}. This spectral data set is publicly available on the ENGRAVE webpage\footnote{\url{www.engrave-eso.org/AT2017gfo-Data-Release}}, and on WISeREP\footnote{\url{https://wiserep.weizmann.ac.il}} \citep{wiserep}.

\par
We also include the +0.5\,d optical spectrum from \cite{Shappee2017} in our analysis (no \xsh\ data were taken on that night). This is a featureless, blue spectrum with no identifiable features, which we model to demonstrate consistency back to this very early epoch. We checked the flux calibration, and verified that it agrees well with the photometry available at this epoch.

\par
We use the \xsh\ set of observed spectra as they are a complete sequence taken daily for ten days, with good signal-to-noise (using an 8\,m aperture telescope), full wavelength coverage across all wavelengths accessible from the ground ($\sim 0.3 - 2.5$\,\micron), and they are well calibrated. While spectra were taken with other telescopes at the time \citep{Andreoni2017, Chornock2017, Kasliwal2017, McCully2017, Nicholl2017, Shappee2017, Smartt2017, Troja2017}, no other instrument provides as complete wavelength coverage, daily temporal coverage, and sensitivity for \gfo. Excellent agreement between the observed photometry and synthetic photometry measured on the flux calibrated \xsh\ spectra (from the ENGRAVE re-calibrations) indicate that this is the definitive spectral data sequence that should be used for modelling \gfo.

\section{Atomic data} \label{sec:Atomic data}
\par
Our modelling work is heavily dependent on accurate and reliable atomic data. As such, we have amassed a comprehensive data set that encompasses the species of interest for our study here. The main data set we use is the default data set that comes with \tardis\ (\textsc{standard}), which is based on Chianti data for H and He \citep[][]{Chianti-OG, Chianti-v10}, and Kurucz for all other elements \citep[][]{Kurucz2017}. This atomic data set contains line information for the lowest few ionisation states for all elements up to the first \rpro\ peak \mbox{($Z \sim 40$)}. Beyond this, it contains some information but it is very incomplete, and is missing many ions of interest for our modelling. This \textsc{standard} data set contains only transitions between known (and laboratory measured) energy levels. Although the wavelengths for all these transitions are reliable, the line list will be incomplete, since it will be missing transitions between levels that have not been experimentally measured. 

\par
Since we are especially interested in \rpro\ species, we added the extended Kurucz atomic data (\url{http://kurucz.harvard.edu/atoms.html}) for \mbox{Sr\,\textsc{i}--\textsc{iii}}, \mbox{Y\,\textsc{i}--\textsc{ii}} and \mbox{Zr\,\textsc{i}--\textsc{iii}} (\textsc{atoms}). These extended line lists contain transitions between both measured and theoretically predicted levels, and so they contain many more lines, albeit with potentially somewhat uncertain wavelength values. Table~\ref{tab:Atomic data lines} illustrates the number of lines that were added to our data set by expanding our list to include the predicted lines for the first \rpro\ peak elements.

\begin{table}
    \centering
    \caption{
        Number of lines belonging to various species and elements of interest. We preferentially included the \textsc{atoms}, \textsc{dream} and \textsc{qub} data over that included within the \textsc{standard} atomic line list for all species and elements listed below. The \textsc{qub} data are described by \citet{McCann2022}.
    }
    \begin{tabular}{ccccc}
        \hline
        \hline
        \multirow{2}{*}{\begin{tabular}[c]{@{}c@{}}Species\end{tabular}}     &\multicolumn{4}{c}{Line list}     \\
        \cline{2-5}
             &\textsc{standard}    &\textsc{atoms}    &\textsc{dream}      &\textsc{qub}    \\
        \hline
        \addlinespace[1ex]
        $^{1}$H -- $^{37}$Rb     &263204     &--     &--     &--    \\
        \addlinespace[1ex]
        $^{38}$Sr\I\ 	 &74 	 &3359      &--    &--     \\
        $^{38}$Sr\II\	 &106 	 &1341      &--    &--     \\
        $^{38}$Sr\III\	 &0 	 &15541     &--    &--     \\
        $^{39}$Y\I\ 	 &328 	 &68263     &--    &--     \\
        $^{39}$Y\II\ 	 &186 	 &43434     &--    &--     \\
        $^{40}$Zr\I\ 	 &693 	 &256457    &--    &--     \\
        $^{40}$Zr\II\ 	 &500 	 &167948    &--    &--     \\
        $^{40}$Zr\III\   &0 	 &38197     &--    &--     \\
        \addlinespace[1ex]
        $^{41}$Nb -- $^{56}$Ba 	 &6772 	 &--     &--     &-- \\
        \addlinespace[1ex]
        $^{57}$La 	 &542 	 &--    &1767     &--   \\
        $^{58}$Ce 	 &2553 	 &--    &16018    &--   \\
        $^{59}$Pr 	 &577 	 &--    &13937    &--   \\
        $^{60}$Nd 	 &1279 	 &--    &3693     &--   \\
        $^{61}$Pm 	 &0 	 &--    &0        &--   \\
        $^{62}$Sm 	 &1577 	 &--    &752      &--   \\
        $^{63}$Eu 	 &489 	 &--    &1277     &--   \\
        $^{64}$Gd 	 &1399 	 &--    &1443     &--   \\
        $^{65}$Tb 	 &107 	 &--    &900      &--   \\
        $^{66}$Dy 	 &1180 	 &--    &2332     &--   \\
        $^{67}$Ho 	 &90 	 &--    &1158     &--   \\
        $^{68}$Er 	 &902 	 &--    &1453     &--   \\
        $^{69}$Tm 	 &752 	 &--    &9946     &--   \\
        $^{70}$Yb 	 &385 	 &--    &7548     &--   \\
        $^{71}$Lu 	 &154 	 &--    &206      &--   \\
        \addlinespace[1ex]
        $^{72}$Hf -- $^{77}$Ir 	 &5277 	 &--     &--     &--     \\
        \addlinespace[1ex]
        $^{78}$Pt\I      &157 	 &--    &--     &1737       \\
        $^{78}$Pt\II 	 &0 	 &--    &--     &14868      \\
        $^{78}$Pt\III 	 &0 	 &--    &--     &65468      \\
        $^{79}$Au\I 	 &61 	 &--    &--     &152        \\
        $^{79}$Au\II 	 &0 	 &--    &--     &2872       \\
        $^{79}$Au\III 	 &0 	 &--    &--     &6792       \\
        \addlinespace[1ex]
        $^{80}$Hg -- $^{88}$Ra 	 &169 	 &--    &--     &--     \\
        \addlinespace[1ex]
        $^{89}$Ac 	 &0 	 &--    &--     &--     \\
        $^{90}$Th 	 &2035 	 &--    &--     &--     \\
        $^{91}$Pa 	 &0 	 &--    &--     &--     \\
        $^{92}$U 	 &1140 	 &--    &--     &--     \\
        \hline
    \end{tabular}
    \label{tab:Atomic data lines}
\end{table}

\par
Due to the complications involved with generating atomic data for heavier elements, with more complex structures beyond \mbox{$Z \sim 40$}, the Kurucz data base is, of course, incomplete. To combat this, we have supplemented these data sets with others that have coverage for some of these heavy species. The \textit{Database on Rare Earths At Mons university} \citep[\textsc{dream},][]{DREAM1, DREAM2} contains atomic line info for some of the lowest ionisation stages of each of the lanthanides, and Lu \mbox{($57 \leq Z \leq 71$)}. This database contains line information for transitions between experimentally measured levels, and so all the lines will have reliable wavelengths.

\par
We ingested all the data available, and preferentially used these values over any line data that were available within the \textsc{standard} data for the same species. Table~\ref{tab:Atomic data lines} contains information on the increased number of lines added by including the \textsc{dream} data. Finally, we also use new data for neutral, singly and doubly ionised platinum and gold, calculated by \cite{McCann2022} and previously used by \cite{Gillanders2021}. The number of lines for these species are also included in Table~\ref{tab:Atomic data lines} (\textsc{qub}).

\par
Using all these atomic data sources, we generated an atomic data set for use by \tardis\ (see Section~\ref{sec:Spectral analysis method}). For this, we used the \carsus\ package\footnote{\url{https://github.com/tardis-sn/carsus}}, which extracted the level energies and statistical weights, and also the Einstein \Aval\ for all transitions. The ionisation information for all species included in our atomic data set are extracted from the National Institute of Standards and Technology Atomic Spectra Database \citep[NIST ASD;][]{NIST2020}.

\section{Composition profiles} \label{sec:Composition profiles}
\par
One of the primary aims of modelling the spectra of \gfo\ is to determine which elemental species are responsible for the observed features. We also want to determine if the overall shape of the spectra, and their evolution, is consistent with compositions expected for BNS mergers. From a hypothetical standpoint, one can imagine undertaking a study where the abundance of every element is treated as a free parameter, with the aim to empirically constrain the composition of the ejecta of \gfo\ at different epochs, in an effort to understand its evolution. However, this approach is not feasible in practice. It would lead to too many free parameters, taking too long to explore, as well as wasting computational resources. Therefore, to save time, we opted to not treat each element as a free parameter, but instead to base our modelling on theoretical work, with compositions obtained from nucleosynthetic calculations from hydrodynamical simulations of BNS mergers. This approach greatly reduces the number of parameters under investigation and provides physically motivated and realistic compositions.

\par
We based our work on the composition extracted from the dynamical mass ejection simulated by a realistic hydrodynamical simulation of a BNS merger \citep[as presented by][]{Goriely2011, Goriely2013, Goriely2015, Bauswein2013}, which we artificially modify to obtain a larger variation in the composition. Specifically, we adopt the hydrodynamical evolution from a smooth particle hydrodynamics simulation of a merger of two stars, both with $M = 1.35$\,\msun, employing the SFHo equation of state \citep{Steiner2013}. Since we use smooth particle hydrodynamics, the simulation provides a number of fluid element trajectories representing the ejecta. For those, we set the initial \Ye\ by hand, which we choose to be proportional to the inclination angle $\theta$ of the given fluid element, to resemble a higher \Ye\ towards the poles. More specifically, we assign $Y_{\rm e} = |\theta| / \pi$, with the inclination angle $\theta$ measured from the equator, such that we obtain $0 \leq Y_{\rm e} \leq 0.5$. Finally, we bin all trajectories by angle; i.e.~effectively by \Ye, where we still distinguish ejection towards the northern and southern hemispheres, which features some degree of statistical fluctuation. This setup does not represent a fully self-consistent description of the merger ejecta, but it provides more flexibility with regards to the composition, and still maintains a certain degree of realism. We also remark that we do not include trajectories of matter becoming unbound on longer timescales, since our simulation ends a few tens of milliseconds after merging. 

\par
We run nuclear network calculations for every trajectory, and sum up the composition mass-averaged within the different bins. The nucleosynthesis calculation starts as soon as the temperature drops below $T = 10^{10}$\,K, and the density is below the neutron-drip density \mbox{($\rho_{\mathrm{drip}} \simeq 4.2 \times 10^{11}$\gcm)}, at which point the initial abundances of heavy nuclei are determined by nuclear statistical equilibrium at the given electron fraction, density and temperature. For the first 10\,ms after merging, the density history is consistently followed by the numerical simulation. Afterwards, the ejected matter is assumed to expand freely with constant velocity. The radii of the ejecta clumps thus grow linearly with time, $t$, and consequently, their densities evolve approximately proportional to $t^{-3}$. As soon as the full reaction network is initiated, the temperature evolution is determined on the basis of the laws of thermodynamics, allowing for possible nuclear heating through $\beta$-decays, fission, and $\alpha$-decays \citep{Meyer1989}.

\par
The nucleosynthesis is followed with a full reaction network, including all 5000 species from protons up to $Z = 110$, lying between the valley of $\beta$-stability and the neutron-drip line \citep[for more details, see][]{Goriely2011, Bauswein2013, Just2015}. All charged-particle fusion reactions on light and medium-mass elements that play a role when the nuclear statistical equilibrium freezes out are included, in addition to radiative neutron captures and photodisintegrations. The reaction rates on light species are taken from the NETGEN library, which includes all the latest compilations of experimentally determined reaction rates \citep{Xu2013}. By default, experimentally unknown reactions are estimated with the TALYS code \citep{Goriely2008} on the basis of the HFB-21 nuclear masses \citep{Goriely2010}. Fission processes, including neutron-induced fission, spontaneous fission, $\beta$-delayed fission, as well as $\beta$-delayed neutron emission, are considered as detailed in \citet{Goriely2015b}. The $\beta$-decay rates are taken from the mean field plus the relativistic QRPA calculation of \citet{Marketin2016}, when not available experimentally. 

\par
\begin{figure*}
    \centering
    \includegraphics[width=1.0\linewidth]{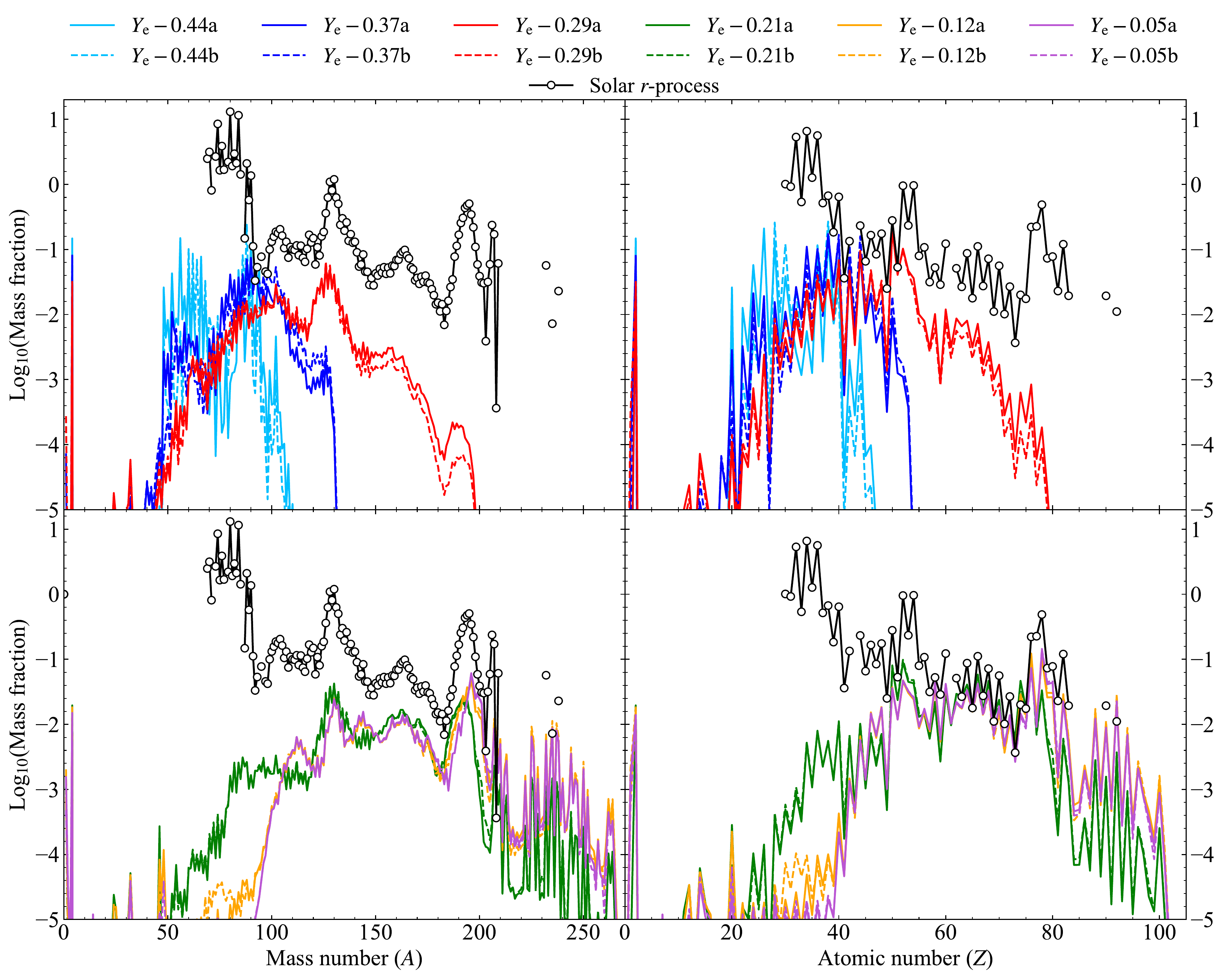}
    \caption{
        Mass fractions of the ejecta material at $t = 1$\,d, as a function of mass number, $A$ (left panels) or atomic number, $Z$ (right panels), divided into the twelve \Ye\ composition bins. The solar system \rpro\ abundance distribution (open circles) from \citet{Goriely1999}, arbitrarily normalised, is shown for comparison.
    }
    \label{fig:Mass fraction compositions plot}
\end{figure*}

The different \Ye\ regimes produce starkly different compositions, as shown in Figure~\ref{fig:Mass fraction compositions plot}, and this is why we consider the compositions via angle/\Ye\ bins; it provides us with a diverse range of composition profiles for use in our radiative transfer modelling. The bins are named corresponding to their average \Ye\ value, and have either the suffix a or b, depending on whether the compositions were extracted from above or below the equatorial plane. In Table~\ref{tab:Ye bins (top 10)}, we present lists of the top ten most abundant elements (as well as a few other elements of interest), and their relative mass fractions, for each of the twelve compositions we have extracted from the simulation. We also present the `Average' composition for reference, which represents the composition averaged across the entire simulation.

\setlength{\tabcolsep}{1.4pt}
\begin{table*}
    \scriptsize
    \centering
    \caption{
        Ten most abundant elements in each of the twelve \Ye\ compositions, plus some other elements of interest. We have also included the abundance averaged across the entire simulation (`Average') for completeness. The final row of the table provides the sum of the lanthanide mass fraction (\Xlanth) in each of the \Ye\ compositions. Column headers: El. -- Element symbol, M.F. -- Mass Fraction. The values are presented in scientific notation, where the bracketed number is the exponent; e.g. $3.13(-1) = 3.13 \times 10^{-1}$. A complete version of this table containing the entire abundance lists is included in the supplementary materials.
    }
    \begin{tabular}{cccccccccccccccccccccccccccccccccccccc}
        \hline
        \hline
        \multicolumn{2}{c}{\AngI}  & &\multicolumn{2}{c}{\AngXII} &   &\multicolumn{2}{c}{\AngII} &   &\multicolumn{2}{c}{\AngXI} &   &\multicolumn{2}{c}{\AngIII} & &\multicolumn{2}{c}{\AngX} & &\multicolumn{2}{c}{\AngIV} &   &\multicolumn{2}{c}{\AngIX}  &   &\multicolumn{2}{c}{\AngV} &   &\multicolumn{2}{c}{\AngVIII} &   &\multicolumn{2}{c}{\AngVI} & &\multicolumn{2}{c}{\AngVII} &    &\multicolumn{2}{c}{Average} \\
        \cline{1-2}
        \cline{4-5}
        \cline{7-8}
        \cline{10-11}
        \cline{13-14}
        \cline{16-17}
        \cline{19-20}
        \cline{22-23}
        \cline{25-26}
        \cline{28-29}
        \cline{31-32}
        \cline{34-35}
        \cline{37-38}
        El.      &M. F. &   &El.      &M. F. &  &El.      &M. F. &  &El.      &M. F. &  &El.      &M. F. &  &El.      &M. F. &  &El.      &M. F. &  &El.      &M. F. &  &El.      &M. F. &  &El.      &M. F. &  &El.      &M. F. &  &El.      &M. F.    &  &El.      &M. F.     \\
        \hline
        Fe  &3.13(-1)  &  &Sr  &3.68(-1)  &  &Sr  &1.94(-1)  &  &Sr  &3.31(-1)  &  &Sn  &2.18(-1)  &  &Sn  &2.26(-1)  &  &Os  &1.10(-1)  &  &He  &1.50(-1)  &  &Os  &1.71(-1)  &  &Pt  &1.53(-1)  &  &Pt  &1.44(-1)  &  &Pt  &1.53(-1)  &  &Pt  &1.23(-1)  \\
        Ni  &1.33(-1)  &  &Ni  &1.93(-1)  &  &Se  &1.81(-1)  &  &Zr  &1.66(-1)  &  &Ru  &1.10(-1)  &  &Te  &1.17(-1)  &  &Te  &9.90(-2)  &  &Te  &7.58(-2)  &  &Pt  &1.61(-1)  &  &Os  &9.77(-2)  &  &Os  &6.79(-2)  &  &Os  &6.62(-2)  &  &Os  &8.00(-2)  \\
        Sr  &1.19(-1)  &  &Kr  &1.45(-1)  &  &Zr  &1.29(-1)  &  &Ru  &1.15(-1)  &  &Te  &9.87(-2)  &  &Ru  &8.51(-2)  &  &Sn  &5.84(-2)  &  &Os  &7.03(-2)  &  &Pb  &4.26(-2)  &  &Pb  &4.75(-2)  &  &Dy  &5.44(-2)  &  &Te  &5.11(-2)  &  &Te  &5.61(-2)  \\
        He  &1.19(-1)  &  &Zn  &9.00(-2)  &  &Kr  &1.17(-1)  &  &He  &1.03(-1)  &  &Zr  &7.22(-2)  &  & I  &5.27(-2)  &  &Pt  &5.37(-2)  &  &Pt  &5.86(-2)  &  &Dy  &3.62(-2)  &  & U  &4.54(-2)  &  &Te  &5.16(-2)  &  &Dy  &4.83(-2)  &  &Dy  &4.53(-2)  \\
        Kr  &8.28(-2)  &  &He  &5.37(-2)  &  &Ru  &6.75(-2)  &  &Mo  &7.03(-2)  &  &Pd  &5.32(-2)  &  &Sb  &4.88(-2)  &  & I  &5.14(-2)  &  &Dy  &4.85(-2)  &  &Te  &3.60(-2)  &  &Te  &4.25(-2)  &  &Ce  &5.06(-2)  &  &Ce  &4.73(-2)  &  &Nd  &4.03(-2)  \\
        Cr  &7.74(-2)  &  &Rb  &3.90(-2)  &  &Mo  &4.70(-2)  &  & Y  &2.92(-2)  &  &Mo  &5.28(-2)  &  &Zr  &4.80(-2)  &  &Dy  &5.01(-2)  &  &Sn  &4.33(-2)  &  &Au  &3.39(-2)  &  &Au  &4.12(-2)  &  &Nd  &5.01(-2)  &  &Nd  &4.72(-2)  &  &Sn  &3.93(-2)  \\
        Zr  &5.77(-2)  &  &Se  &2.26(-2)  &  &He  &4.58(-2)  &  &Pd  &2.92(-2)  &  & I  &4.35(-2)  &  &Pd  &4.73(-2)  &  &Er  &4.13(-2)  &  & I  &4.14(-2)  &  & U  &3.13(-2)  &  &Dy  &3.75(-2)  &  &Er  &4.53(-2)  &  &Hg  &4.44(-2)  &  &Ce  &3.91(-2)  \\
        Zn  &2.59(-2)  &  &Ge  &2.11(-2)  &  &Rb  &3.14(-2)  &  &Se  &2.67(-2)  &  &Sb  &4.25(-2)  &  &Xe  &3.56(-2)  &  &Xe  &4.10(-2)  &  &Er  &4.05(-2)  &  &Nd  &3.11(-2)  &  &Ce  &3.61(-2)  &  &Gd  &4.01(-2)  &  &Au  &4.41(-2)  &  &Er  &3.71(-2)  \\
        Ca  &1.50(-2)  &  &Zr  &1.66(-2)  &  & Y  &3.08(-2)  &  &Kr  &2.43(-2)  &  &Cd  &3.94(-2)  &  &Cd  &3.52(-2)  &  & W  &3.51(-2)  &  &Nd  &4.04(-2)  &  &Er  &3.08(-2)  &  &Nd  &3.31(-2)  &  &Au  &3.68(-2)  &  &Er  &3.99(-2)  &  & I  &3.59(-2)  \\
         Y  &8.75(-3)  &  &Fe  &1.08(-2)  &  &Br  &2.52(-2)  &  &Rh  &1.05(-2)  &  &Sr  &3.25(-2)  &  &He  &3.45(-2)  &  &Gd  &3.49(-2)  &  &Xe  &3.71(-2)  &  &Ce  &2.99(-2)  &  &Hg  &3.29(-2)  &  & I  &3.67(-2)  &  & I  &3.57(-2)  &  &Gd  &3.34(-2)  \\
        \addlinespace[1ex]
        Nd  &6.05(-5)  &  &Y   &4.72(-3)  &  &Ce  &1.58(-5)  &  &Ce  &4.95(-6)  &  &Nd  &1.03(-2)  &  &Sr  &2.47(-2)  &  &Nd  &2.70(-2)  &  &Ce  &3.20(-2)  &  &Sm  &1.67(-2)  &  &Sm  &1.87(-2)  &  &Sm  &2.44(-2)  &  &Sm  &2.23(-2)  &  &Sm  &2.16(-2)  \\
        Ce  &6.02(-5)  &  &Nd  &4.86(-5)  &  &Nd  &1.48(-5)  &  &Nd  &2.60(-6)  &  &Ce  &9.64(-3)  &  &Nd  &1.03(-2)  &  &Sm  &2.60(-2)  &  &Sm  &2.61(-2)  &  &Eu  &1.33(-2)  &  &Eu  &1.55(-2)  &  &Eu  &1.96(-2)  &  &Eu  &1.77(-2)  &  &Eu  &1.74(-2)  \\
        Sm  &4.45(-5)  &  &Sm  &4.64(-5)  &  &Sm  &7.79(-6)  &  &Sm  &1.55(-6)  &  &Y   &7.96(-3)  &  &Sm  &8.45(-3)  &  &Eu  &2.20(-2)  &  &Eu  &2.06(-2)  &  &Zr  &2.69(-4)  &  &Zr  &3.31(-4)  &  &Zr  &1.67(-4)  &  &Zr  &1.72(-4)  &  &Sr  &8.56(-3)  \\
        Eu  &4.04(-5)  &  &Eu  &3.83(-5)  &  &Eu  &6.14(-6)  &  &Eu  &1.07(-6)  &  &Sm  &6.40(-3)  &  &Ce  &7.98(-3)  &  &Ce  &1.83(-2)  &  &Zr  &4.04(-3)  &  &Sr  &5.31(-5)  &  &Sr  &4.38(-5)  &  &Sr  &1.60(-5)  &  &Sr  &1.88(-5)  &  &Zr  &7.19(-3)  \\
        --  &--        &  &Ce  &3.44(-5)  &  &--  &--        &  &--  &--        &  &Eu  &4.59(-3)  &  &Eu  &6.35(-3)  &  &Zr  &1.08(-2)  &  &Sr  &2.03(-3)  &  &Y   &1.95(-5)  &  &Y   &1.83(-5)  &  &Y   &7.00(-6)  &  &Y   &7.35(-6)  &  &Y   &1.07(-3)  \\
        --  &--        &  &--  &--        &  &--  &--        &  &--  &--        &  &--  &--        &  &Y   &6.00(-3)  &  &Sr  &6.47(-3)  &  &Y   &4.91(-4)  &  &--  &--        &  &--  &--        &  &--  &--        &  &--  &--        &  &--  &--        \\
        --  &--        &  &--  &--        &  &--  &--        &  &--  &--        &  &--  &--        &  &--  &--        &  &Y   &1.46(-3)  &  &--  &--        &  &--  &--        &  &--  &--        &  &--  &--        &  &--  &--        &  &--  &--        \\
        \addlinespace[1ex]
        \Xlanth\  &5.69(-4)  &  &\Xlanth\  &5.14(-4)  &  &\Xlanth\  &1.15(-4)  &  &\Xlanth\  &2.50(-5)  &  &\Xlanth\  &4.99(-2)  &  &\Xlanth\  &6.90(-2)  &  &\Xlanth\  &2.99(-1)  &  &\Xlanth\  &3.29(-1)  &  &\Xlanth\  &2.44(-1)  &  &\Xlanth\  &2.58(-1)  &  &\Xlanth\  &3.79(-1)  &  &\Xlanth\  &3.42(-1)  &  &\Xlanth\  &3.12(-1)  \\
        \hline
    \end{tabular}
    \label{tab:Ye bins (top 10)}
\end{table*}
\setlength{\tabcolsep}{6pt}

\par
These different compositions provide `representative' compositions that are feasible for BNS mergers. However, we wish to highlight the following points. First, since the bulk of the material is ejected along the equatorial plane, and the nucleosynthesis calculations are based on a sample of massive tracer particles, the ejecta are generally better sampled for equatorial bins; i.e.~lower \Ye. As we move towards the poles, the amount of material (and thus number of simulated nucleosynthesis tracer particles) in these angle bins drops off significantly, leading to more poorly sampled compositions, because of the significantly smaller solid angle. This leads to some larger scatter within the composition of these bins, which will get worse for bins with higher \Ye. This becomes most apparent when comparing our high \Ye\ compositions, \AngI\ and \AngXII. These composition bins should be comparable, yet the iron abundance varies significantly (0.313 in \AngI, versus 0.0108 in \AngXII).

\par
Second, we wish to clarify that we are \textit{not} using these compositions to represent viewing angle. We have extracted these compositions in this manner to provide us with a diverse range, or selection, of composition profiles that could feasibly be produced from a BNS merger, as opposed to summing up the entire ejecta from the simulation, which would provide a single composition. This set of compositions are what we base our sequence of spectral models upon.

\par
Finally, it should be emphasised that the thermodynamic properties of each trajectory is given by the hydrodynamical simulation. In particular, each of the twelve \Ye\ bins is found to be ejected with a rather similar average expansion velocity, $\langle v \rangle \simeq 0.25 \pm 0.05 \, c$, except along the pole, where larger average velocities $\sim 0.4 \, c$ are found. Similarly, each of the twelve \Ye\ bins are characterised by initial entropies per nucleon of $\sim 25 \pm 5 \, \rm{k_{\textsc{b}}}$. For this reason, in each bin, the electron fraction is the key property governing the nucleosynthesis.

\section{Spectral analysis method} \label{sec:Spectral analysis method}
\par
Numerous attempts have been made to interpret the early spectra of \gfo. Many of these have invoked the two-component model (rapidly expanding, low opacity, blue component, followed by a slower moving, high opacity, red component) to explain the evolution of the early SEDs \citep{Chornock2017, Cowperthwaite2017, Kasen2017, Coughlin2018}. However, other works dispute multi-component models being necessary to explain the early evolution of \gfo. For example, \cite{Smartt2017} and \cite{Waxman2018} argue that the lightcurve and spectra can be adequately reproduced with a single ejecta component, with low-to-moderate opacity. Throughout this work, we use single zone models to calculate theoretical spectra of \gfo. Using different compositions from the hydrodynamical simulations allows us to test if a single component, or multiple, are required to explain the observed spectral evolution.

\par
For all the modelling we present in this work, we use the 1D Monte Carlo radiative transfer code, \tardis\ \citep[\textit{Temperature And Radiative Diffusion In Supernovae};][]{tardis, tardis2}. This code is capable of rapidly generating synthetic spectra of explosive transients. \tardis\ begins by assuming a spherically symmetric explosion, with ejecta material that is expanding homologously. The code assumes an inner boundary (the properties of which are determined from user inputs), beneath which the ejecta material is completely optically thick. The line-forming region is the expanding material above this boundary. \tardis\ assumes that the radiation emerging from this boundary perfectly resembles a blackbody. This inner boundary, or `photospheric', approximation greatly reduces the complexity of the simulation, since we do not consider the internal powering mechanism. It also reduces the velocity space we need to simulate, since the code need only simulate the ejecta properties above the inner boundary. This assumption of a sharp photosphere within our \tardis\ models can impact the results inferred from modelling. In a real astrophysical explosion, there is no single position where the material transitions between being optically thick and thin. In reality, this position will vary with wavelength, and so the position of this sharp inner boundary within \tardis\ is unphysical, and affects the model SED.

\par
One would expect to be able to `see' deeper within the ejecta in the NIR than in the optical or the UV. Therefore, if the photosphere is positioned such that it agrees with the data in the UV and optical parts of the spectrum, then we will lose information from NIR photons beneath this boundary. This will lead to disagreements with data at NIR wavelengths. To combat this problem, the inner boundary position would need to either be computed in a wavelength-dependent manner, or removed entirely, both of which are more computationally expensive and not possible with the current version of \tardis. Despite this restriction, \tardis\ can still reproduce the SEDs of observed transients well (at least in the UV and optical parts of the spectrum, which are typically the focus for spectroscopic studies of SNe).

\par
At the beginning of the simulation, \tardis\ generates $r$-packets, which represent bundles of photons emerging from the optically thick photosphere in the model explosion. The $r$-packets are randomly assigned frequencies based on the temperature at the inner boundary. The simulation begins, and these $r$-packets are then free to randomly propagate through the expanding outer regions of ejecta, the properties of which (e.g. density, temperature and composition), are determined by user inputs. As the $r$-packets propagate, they are free to interact with the expanding ejecta material, and these interactions\footnote{Throughout this paper, where we mention the term `interaction', we are referring to the photon--ion/atom interactions within our \tardis\ simulations, unless otherwise stated.} lead to discernible features in the computed spectrum (by free electron scattering or absorption/re-emission in bound-bound and bound-free transitions).

\par
When the simulation ends, all $r$-packets that escaped the outer boundary are used to compute a synthetic spectrum. \tardis\ is time-independent, which means that it can only generate model spectra for a single epoch. However, by modelling spectra captured across multiple epochs, and evolving the model input parameters in a consistent way with time, one can generate a physically motivated sequence of model spectra that can be used to understand the evolution of the explosive transient under investigation, across some, or most, of its observed spectral sequence (provided the object remains in a photospheric regime).

\par
\begin{figure}
    \centering
    \includegraphics[width=\linewidth]{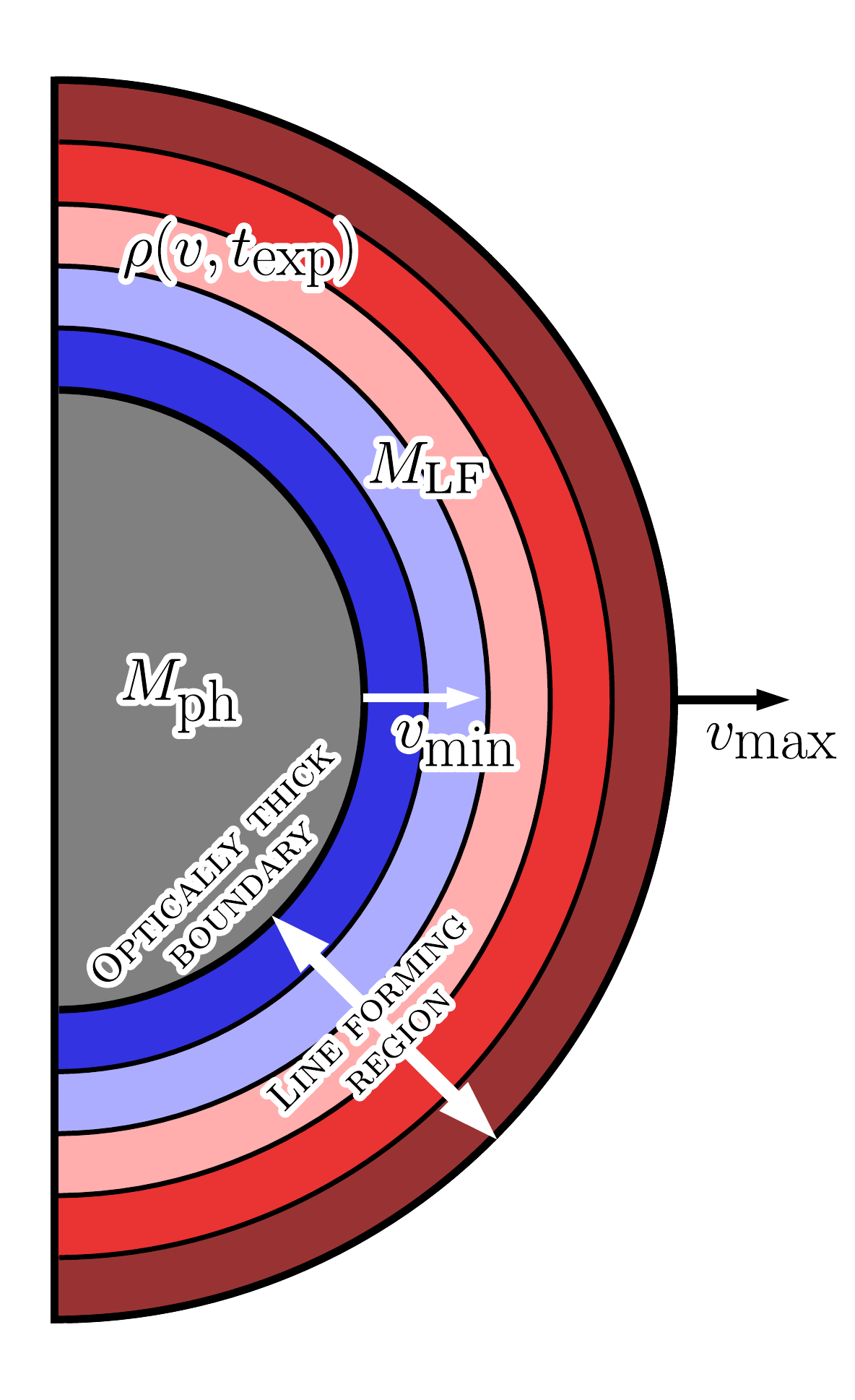}
    \caption{
        Schematic illustrating how \tardis\ functions.
    }
    \label{fig:TARDIS schematic}
\end{figure}

Figure~\ref{fig:TARDIS schematic} schematically illustrates how \tardis\ functions, with the optically thick inner boundary (or photosphere) at velocity $v_{\rm min}$ and temperature $T$, producing a blackbody continuum spectrum. The line-forming region is where the $r$-packets interact with the expanding ejecta material. The density profile, $\rho(v, \, t_{\rm exp})$, controls the total mass of material above the photosphere. \tardis\ is capable of simulating a layered or stratified ejecta. We note that we are not sensitive to any mass beneath the photosphere (denoted $M_{\rm ph}$) and the masses we refer to throughout this paper will correspond to the mass enclosed by the line-forming region within our models, $M_{\textsc{lf}}$. The total ejecta mass of the system is $M_{\rm ej}= M_{\rm ph} + M_{\textsc{lf}}$, and typically $M_{\textsc{lf}} \ll M_{\rm ph}$. Although we do not constrain the total mass of the system, we are able to constrain the composition of the ejecta where our observations are most sensitive (the line-forming region), which allows us to deduce the relative mass fractions of the species of interest in the ejecta of \gfo.

\par
We note that \tardis\ has been used previously to model the early spectra of \gfo\ \citep{Smartt2017, Watson2019, Gillanders2021, Perego2022}. Since KNe have typical ejecta speeds, $v \gtrsim 0.1\,c$ at early times \citep{Bauswein2013, Hotokezaka2013, Siegel2017, Ciolfi2017, Radice2018}, relativistic effects should be accounted for in the radiative transfer. Therefore, we use the full treatment of special relativity within \tardis, developed by \cite{Vogl2019}, throughout all of our analysis presented here. We use the \texttt{LTE} (local thermal equilibrium) approximation for ionisation, and \texttt{dilute-LTE} for excitation. We use the most sophisticated line interaction treatment, \texttt{macroatom}, which provides an accurate representation of fluorescence and multi-line effects. For all our models we use a power law density profile, of the general form:
\begin{equation}
    \rho(v, t_{\rm exp}) = \rho_{\rm 0}\left(\frac{t_{\rm 0}}{t_{\rm exp}}\right)^{3}\left(\frac{v}{v_{\rm 0}}\right)^{-\Gamma}
    \label{eqn:Density profile}
\end{equation}
for ${v_{\rm min}} \leq v \leq {v_{\rm max}}$, where $\rho_{\rm 0}$, $t_{\rm 0}$, $v_{\rm 0}$, $\Gamma$ and ${v_{\rm max}}$ are constants. We chose values for these constants empirically to reproduce the SED of the early spectra of \gfo. For all our models, we use \mbox{$t_{\rm 0} = 2$ days} and \mbox{$v_{\rm 0} = 14000$\,\kms}. $\rho_0$ essentially controls the amount of material present in the model, and for our modelling approach in this work we initially treat $\rho_0$ as a free parameter, in order to obtain reasonable fits for each composition profile across all epochs\footnote{Note that we find consistency across our sequence of best-fitting models, with constant values for $\rho_0$ (and also $t_0$ and $v_0$) producing a consistent density profile.}. We use an exponent, \mbox{$\Gamma = 3$}, as it agrees with hydrodynamical simulations of BNS mergers \citep{Hotokezaka2013, Tanaka2013} and has been used with reasonable success in other works modelling \gfo\ \citep{Watson2019, Gillanders2021}. We use $v_{\rm max} = 0.35\,c$ as the maximum ejecta velocity for our models, as there is no spectroscopic evidence for material at velocities higher than this. In its standard LTE mode of operation, \tardis\ estimates the temperature profile using a blackbody model for the local radiation field. Here, however, we are cautious of using this approach, given the very complex (partly unknown) distribution of opacities, and the extreme line-blanketing at UV and blue wavelengths that are expected for ejecta rich in heavy elements. Since these effects make a local blackbody radiation field mode questionable, we opt instead to simply adopt a fixed temperature throughout the entire ejecta for each of our \tardis\ models. This avoids introducing a complex (and likely inaccurate) temperature profile in the models, and makes it clearer how the model parameters control the shape of the SED and ionisation of the ejecta material. Although a fixed temperature throughout our model ejecta is simplistic, we note that our largest systematic uncertainty does not arise from this, but is due to the incompleteness of the atomic data and the simple excitation/ionisation assumptions. Our model temperatures at each epoch were deduced by fitting a \tardis\ model continuum to the observed spectrum, allowing some minor deviation if it improved the overall fit to observations, when the effects of line interactions were considered. In this approach, our model luminosity is not a free parameter, and is computed via
\begin{equation}
    L = 4 \pi \left( v_{\rm min} \times t_{\rm exp} \right)^{2} \sigma_{\textsc{r}} T^{4}
    \label{eqn:Luminosity/Temperature relationship}
\end{equation}
assuming homologous expansion, where $\sigma_{\textsc{r}}$ is the Stefan-Boltzmann constant and $T$ is the model temperature. Finally, $t_{\rm exp}$ was well-constrained at all epochs from the GW detection associated with \gfo\ \citep{LigoVirgo2017}.

\par
\begin{figure}
    \centering
    \includegraphics[width=\linewidth]{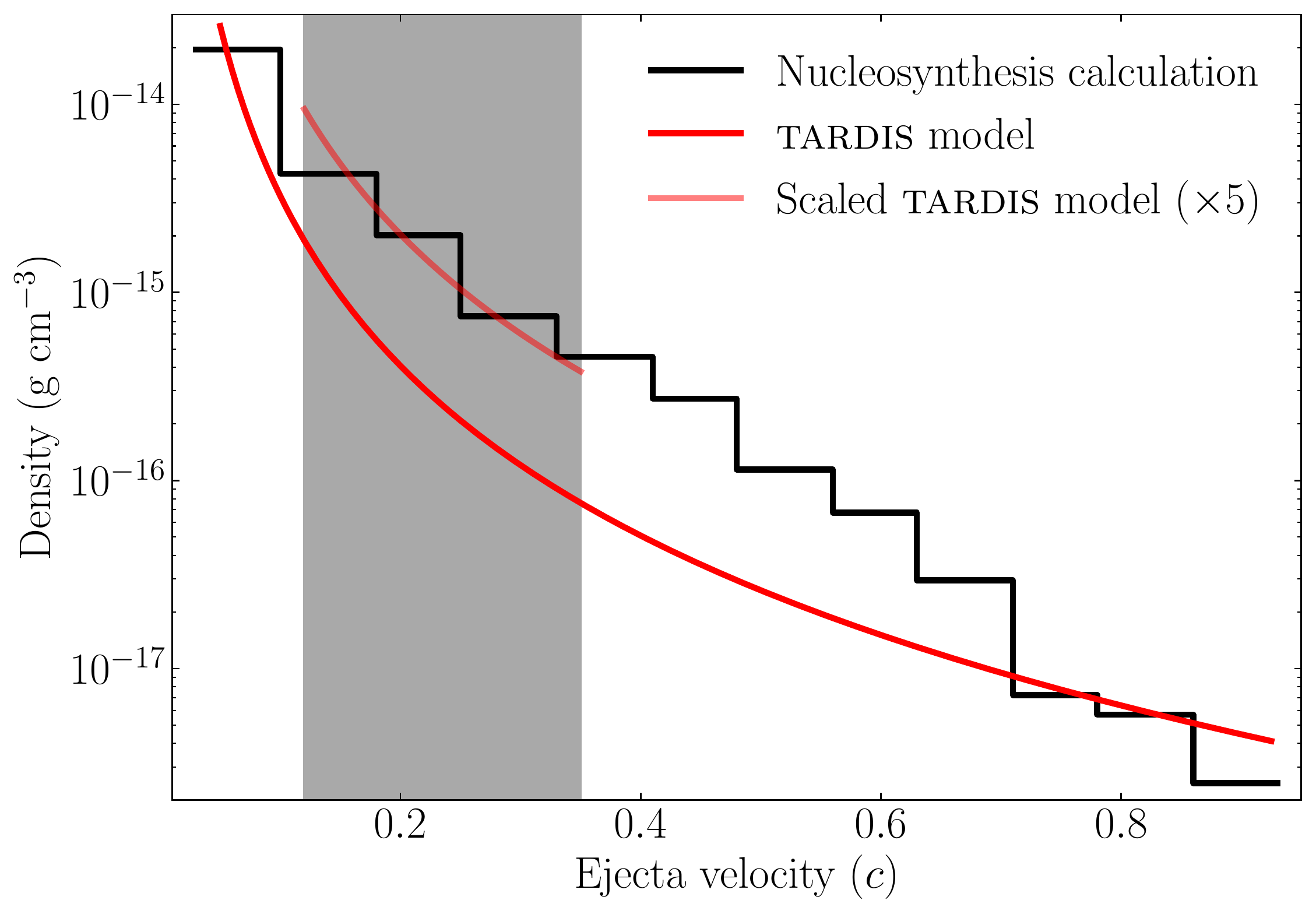}
    \caption{
        Comparison of the density profile from our nucleosynthesis calculation, and the density profile used in our \tardis\ modelling. The densities have been computed assuming homologous expansion, and a time since explosion, $t_{\rm exp} = 1$ day. The grey shaded region highlights the velocity range that we are sensitive to in our best-fitting \tardis\ models for the $+1.4 - 4.4$\,d spectra, that we present in Section~\ref{sec:Spectral analysis results}. We also plot our \tardis\ density profile scaled by a constant factor ($\times 5$) within the velocity range we explore with our \tardis\ models. This illustrates how the distribution of ejecta material across this velocity range ($0.12 - 0.35 \, c$) in our \tardis\ models is comparable to that predicted from the hydrodynamic simulation, albeit with less total material included.
    }
    \label{fig:Density profiles}
\end{figure}

As a check, we compare the density profile from the hydrodynamic simulation used in the nucleosynthesis calculations (Section~\ref{sec:Composition profiles}) with the analytic density profile formulated in Equation~\ref{eqn:Density profile}. The hydrodynamical simulation was binned into twelve discrete velocity bins, which span a large velocity range ($0.03 - 0.93 \, c$). Assuming homologous expansion, the corresponding densities for each bin have been derived at 1 day post-explosion. The \tardis\ density has been computed for the same epoch and we compare them in Figure~\ref{fig:Density profiles}. The grey shaded region encompasses the velocity space we are sensitive to in our best-fitting \tardis\ models for the \mbox{$+1.4 - 4.4$\,d} spectra of \gfo, that we present in Section~\ref{sec:Spectral analysis results}. Our \tardis\ model density is a factor of $\sim 5$ lower than in the hydrodynamic simulation, although the exponents of both profiles are comparable. This discrepancy is clearly evident in Figure~\ref{fig:Density profiles}, where the \tardis\ model density has been scaled by a factor of 5 (within the velocity range our \tardis\ models are sensitive to). We see that this resembles the density profile extracted from the hydrodynamic simulation. The discrepancy between our \tardis\ and hydrodynamic simulation density profiles arises from our method of selecting the density profile in \tardis. We treated the \tardis\ density profile as a free parameter, and converged towards the solution that is best able to reproduce the observations across a range of epochs (see Section~\ref{sec:Spectral analysis results}). From this discrepancy, it is apparent that we require less ejecta material in the \tardis\ line-forming region to reproduce the observations than the hydrodynamic simulation predicts. A more in-depth analysis exploring this discrepancy should be undertaken, but is outside the scope of this paper.

\par
When discussing our model results in Section~\ref{sec:Spectral analysis results}, we will explore the \textit{Spectral element DEComposition} (SDEC) plots for our models. These plots are produced by the \tardis\ code to illustrate the contribution each interaction type has on our model spectrum. The SDEC plot assists in highlighting the last interaction the escaping packets have, which allows us to determine what component of the model ejecta has the most prominent effect on our synthetic spectra. Absorption is illustrated by the coloured regions beneath the Flux = 0 level, and the strength of the absorption feature is proportional to its size. Different colours correspond to different ions in the model, with only the most prominent ones highlighted, for clarity. Emission from the last interaction is indicated above the Flux = 0 level, with free electron scattering and the inner boundary also contributing to our emergent spectrum. These plots are vital for highlighting the contribution different species have on our emergent spectrum.

\par
When presenting our results in Section~\ref{sec:Spectral analysis results}, we refer to our `best-fitting' \tardis\ models. All our model comparisons have been performed using a `$\chi$-by-eye' approach, as is commonly performed in SN studies \citep[see e.g.][]{Stehle2005}. This approach suffers from the issue that it can be difficult to determine the validity of similar models, leading to some level of subjectivity. As such, it can be difficult to accurately constrain parameters that only mildly impact the overall model. However, for the model parameters to which our results are primarily sensitive (e.g. temperature, \SrII\ mass) these uncertainties are minimal.

\section{Spectral analysis results} \label{sec:Spectral analysis results}
\par
Here we present the results of our \tardis\ modelling for the spectra of \gfo\ during its earliest epochs. As the transient exhibits rapid spectroscopic evolution, \gfo\ may remain in a photospheric regime only for the first few days post-explosion. Beyond this, the single-temperature blackbody photospheric approximation used within \tardis\ may not be capable of reproducing the spectra in a physically meaningful way. We attempt to replicate the spectra taken within the first week after explosion, but caution that the models for the later epochs (beyond $\sim 4$ days) may not be reliable. All epochs and spectra are referred to by the time from the gravitational wave merger time for GW170817. 

\subsection{Epoch 0: +0.5 day spectrum} \label{sec:Spectral analysis results - Epoch 0}
\par
\begin{figure}
    \centering
    \includegraphics[width=\linewidth]{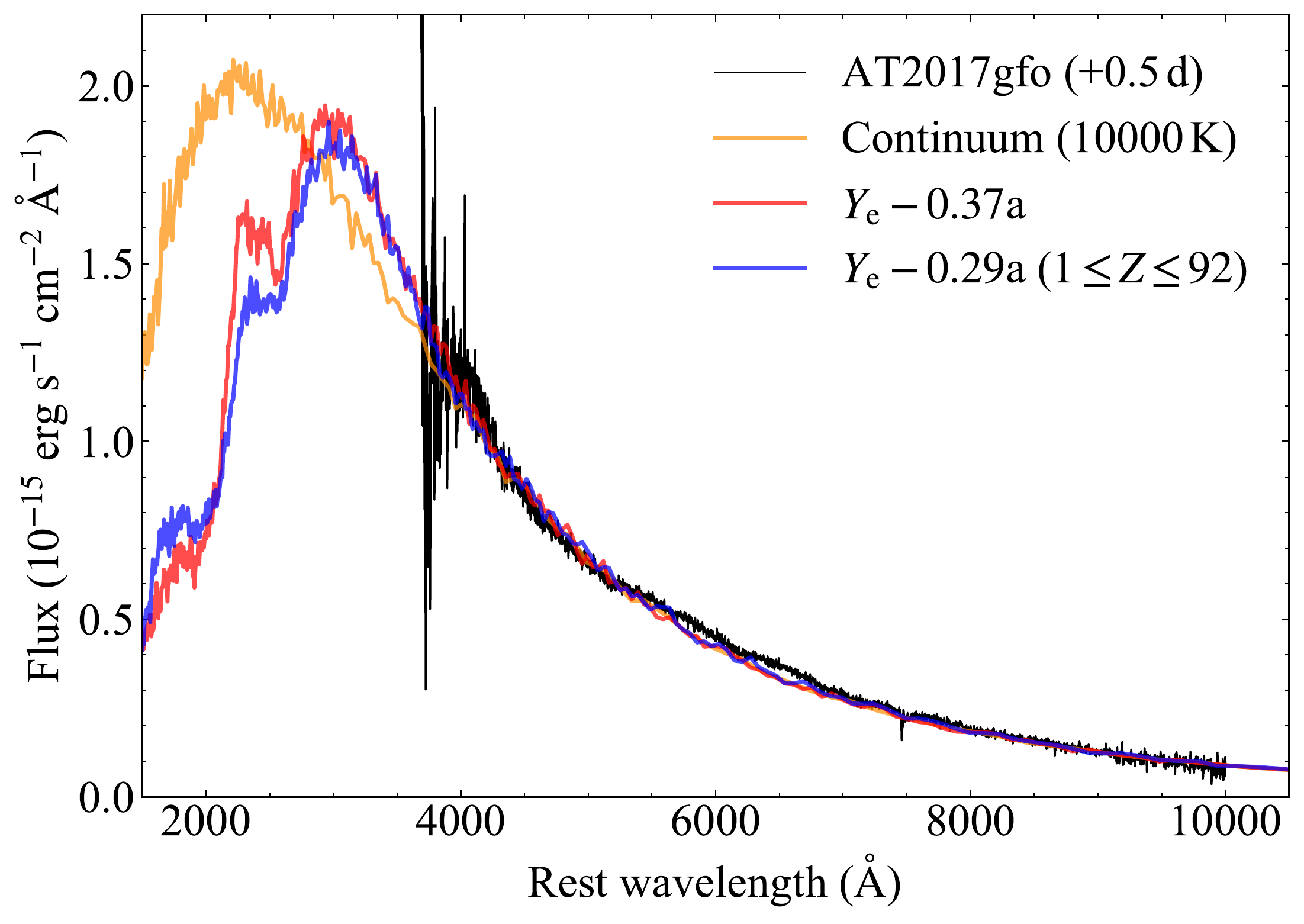}
    \caption{
        Spectrum of \gfo, taken 0.5\,d after the GW trigger, plotted alongside our \AngIII\ model (blue), \AngII\ model (red), and the \tardis\ model continuum, at $T = 10000$\,K (orange).
    }
    \label{fig:0.5d Ang02+03 models}
\end{figure}

This spectrum from \cite{Shappee2017} is the earliest available, taken just 0.5 days after the gravitational wave detection associated with \gfo. We show the spectrum in Figure~\ref{fig:0.5d Ang02+03 models} and it appears to be hot, blue and featureless. We find a good fit to the observations with a \tardis\ continuum temperature, $T = 10000$\,K, which is also plotted. Beyond constraining the temperature at this epoch, we cannot deduce much else about the transient. We cannot directly constrain the composition, due to the lack of observed spectroscopic features. For completeness, we take our best-fitting composition profiles for the subsequent spectra (\AngII\ for the +1.4\,d \xsh\ spectrum, and \AngIII\ for the $+2.4 - 7.4$\,d spectra; see Sections~$\ref{sec:Spectral analysis results - Epoch 1} - \ref{sec:Spectral analysis results - Epochs 5-7}$), and evolve them backwards to this epoch to see if we should expect to see any observable features at this epoch. These models are also plotted in Figure~\ref{fig:0.5d Ang02+03 models}. The models that we present are identical, apart from their composition. For both, we use \mbox{$T = 10000$\,K}, \mbox{$\rho_0 = 1.2 \times 10^{-14}$\gcm}, and \mbox{$v_{\rm min} = 0.30\,c$}. Our \tardis\ model spectra exhibit no strong absorption features at the wavelengths covered by the data, as the ejecta are much too hot. The absence of any absorption features is in agreement with the data. We note that our \tardis\ models exhibit strong absorption at ultra-violet (UV) wavelengths \mbox{($\lambda \lesssim 3000$\,\AA)}, indicating that UV observations of future events are likely to be required if one wants to constrain the ejecta composition during the first $12 - 24$ hours. It will be challenging to confidently identify a counterpart \textit{and} to trigger and acquire high quality spectra from one of the \textit{Hubble Space Telescope's} UV spectrometers (COS or STIS) within the first day of observation of a future kilonova.

\subsection{Epoch 1: +1.4 day spectrum} \label{sec:Spectral analysis results - Epoch 1}
\subsubsection{Best-fitting \Ye\ composition} \label{sec:Spectral analysis results - Epoch 1 - Best-fitting Ye composition}
\par
The spectral evolution of \gfo\ from a hot, featureless blackbody, to a redder spectrum peaking between $0.7 - 1.1\,\mu$m within 2.4 days is unlike any previously observed extragalactic transient. The first \xsh\ spectrum, obtained at +1.4\,d, is cooler than the spectrum at +0.5\,d, but still much bluer than all subsequent spectra. This spectrum almost certainly represents the transient while it is in a photospheric regime, making it a good candidate for accurate \tardis\ modelling \citep[as previously demonstrated by][]{Smartt2017, Watson2019, Gillanders2021, Perego2022}.

\par
The SED of the observed spectrum can be reasonably well-approximated by a \tardis\ model continuum with $T = 4500$\,K. Using the composition profiles discussed in Section~\ref{sec:Composition profiles}, we initially find that the best-fitting composition for our \tardis\ model at this epoch is \AngII, which is dominated by elements from the first \rpro\ peak (see Table~\ref{tab:Ye bins (top 10)}). The broad absorption feature present in the spectrum between \mbox{$\sim 7000 - 10000$\,\AA} can be reproduced well by strontium, which is the dominant element by mass fraction in this \AngII\ composition. Specifically, the absorption feature is caused by the \SrII\ $4\rm{d} - 5\rm{p}$ triplet. This is not a new result; \cite{Watson2019} previously identified \SrII\ as a viable candidate for reproducing this feature in the same \xsh\ spectrum of \gfo, also using \tardis.

\par
Assuming this feature is produced predominantly by \SrII\ absorption, we are able to replicate the feature shape with a model minimum ejecta velocity, \mbox{$v_{\rm{min}} = 0.28\,c$}. To get the model absorption between \mbox{$\sim 7000 - 10000$\,\AA} to match observation (while keeping the composition ratios fixed), we require \mbox{$\rho_0 = 1.2 \times 10^{-14}$\gcm}, corresponding to a \SrII\ model mass (in the \tardis\ line-forming region) of \mbox{\MSrII\ $= 1.6 \times 10^{-7}$\,\msun}. Table~\ref{tab:TARDIS model parameters} contains all relevant model input parameters, as well as information on the total mass in the \tardis\ model, and masses for some of the dominant elements and ions. We highlight again that all masses quoted here, and throughout this manuscript, represent only the mass within the computational domain of the \tardis\ models, illustrated as the line-forming region in Figure~\ref{fig:TARDIS schematic}. From \tardis\ modelling, we cannot independently constrain either the composition or total mass beneath the inner boundary. That is best estimated from the evolution of total luminosity, since that measures total radioactive powering and integrated opacity. \citep[e.g.][]{Smartt2017, Villar2017, Waxman2018, Nicholl2021}. 

\par
\begin{figure}
    \centering
    \includegraphics[width=\linewidth]{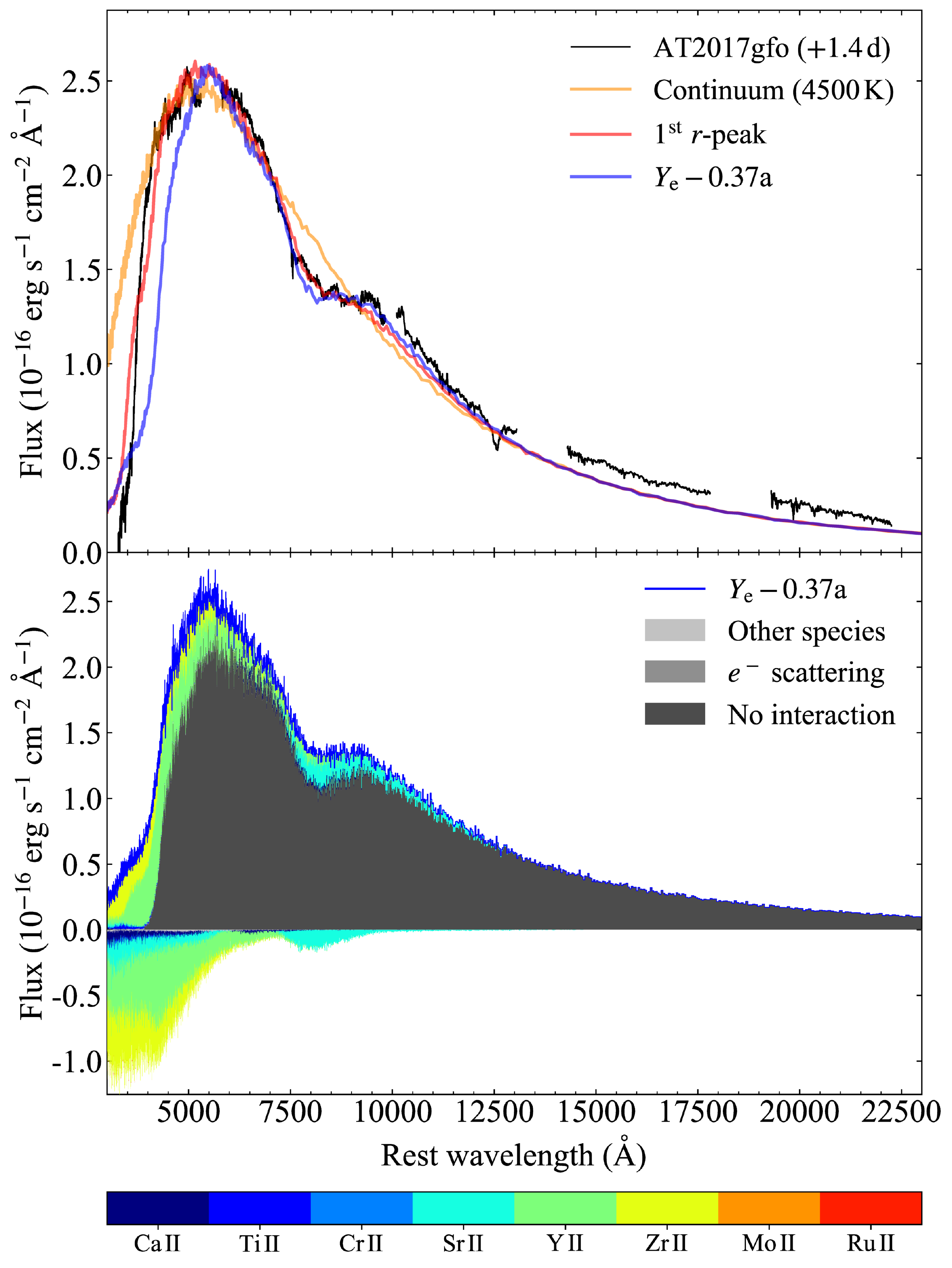}
    \caption{
        \textit{Top panel:} \xsh\ spectrum of \gfo, taken +1.4\,d after the GW trigger, plotted alongside the best-fitting \AngII\ model (blue), the continuum from our best-fitting model (orange), and a model containing only first \rpro\ peak elements (red). \textit{Bottom panel:} SDEC plot for the \AngII\ model, illustrating that \SrII\ produces the broad absorption at $\sim 8000$\,\AA. \YII\ and \ZrII\ produce significant absorption below $\sim 5000$\,\AA.  
    }
    \label{fig:1.4d Ang02 model}
\end{figure}

Our best-fitting \AngII\ model spectrum is shown in the top panel of Figure~\ref{fig:1.4d Ang02 model}. To highlight the effects of radiation transport though the expanding material, we also plot the \tardis\ model continuum; i.e. the input spectrum from the inner boundary before photon packet interaction occurs in the line-forming region. The differences between this continuum model and our best fitting \tardis\ model are due to photon--ion/atom interactions, and so they highlight which regions of the spectrum are most affected by the expanding material.

\begin{table*}
    \centering
    \caption{
        Input model parameters and inferred masses in the line-forming regions of the \tardis\ models for the first four epochs, which we are reasonably confident are in a photospheric regime. The masses present in this region for relevant species are also listed. The model parameters are illustrated and represented in Figure~\ref{fig:TARDIS schematic}. The model parameters in bold correspond to our preferred best-fitting model for each epoch.
    }
    \begin{tabular}{ccccccccc}
        \hline
        \hline
        {\begin{tabular}[c]{@{}c@{}}Epoch \\ (days)\end{tabular}}
        &{\begin{tabular}[c]{@{}c@{}}$v_{\rm min}$ \\ ($c$)\end{tabular}}
        &{\begin{tabular}[c]{@{}c@{}}$T$ \\ (K)\end{tabular}}
        &{\begin{tabular}[c]{@{}c@{}}$\rho_{0}$ \\ ($10^{-15}$\gcm)\end{tabular}}
        &{\begin{tabular}[c]{@{}c@{}}Composition \\ profile\end{tabular}}
        &{\begin{tabular}[c]{@{}c@{}}$M_{\textsc{lf}}$ \\ ($10^{-6}$\,\msun)\end{tabular}}
        &{\begin{tabular}[c]{@{}c@{}}\SrII\ (Sr) mass \\ ($10^{-6}$\,\msun)\end{tabular}}
        &{\begin{tabular}[c]{@{}c@{}}\YII\ (Y) mass \\ ($10^{-6}$\,\msun)\end{tabular}}
        &{\begin{tabular}[c]{@{}c@{}}\ZrII\ (Zr) mass \\ ($10^{-6}$\,\msun)\end{tabular}}   \\
        \hline
        1.4         &0.28           &4500           &12         &\Ye--0.37a                                             &240            &0.16 (47)         &0.83 (7.4)     &14 (31)         \\
        {\bf 1.4}   &{\bf 0.28}     &{\bf 4500}     &{\bf 4.0}  &$\boldsymbol{1^{\rm st}}$ $\boldsymbol{r}$\bf{-peak}   &{\bf 80}       &{\bf 0.10 (60)}   &{\bf --}       &{\bf 5.7 (20)}  \\
        1.4         &0.28           &4500           &12         &\AngIII\ ($1 \leq Z \leq 92$)                          &240            &0.02 (7.8)        &0.16 (1.9)     &6.5 (17)        \\
        1.4         &0.28           &4500           &12         &\AngIII\ ($1 \leq Z \leq 56$)                          &240            &0.02 (7.8)        &0.16 (1.9)     &6.5 (17)        \\
        \addlinespace[1ex]
        {\bf 2.4}   &{\bf 0.20}     &{\bf 3600}     &{\bf 4.0}  &{$\boldsymbol{Y_{\rm e} - 0.29}$\textbf{a} ($\boldsymbol{1 \leq Z \leq 92}$)}  &{\bf 200}  &{\bf 1.9 (6.5)}     &{\bf 1.5 (1.6)}        &{\bf 14 (15)}     \\
        2.4         &0.20           &3600           &4.0        &\AngIII\ ($1 \leq Z \leq 56$)                                                  &200        &1.9 (6.5)           &1.5 (1.6)              &14 (15)           \\
        2.4         &0.20           &3600           &1.0        &\AngII\                                                                        &50         &1.2 (9.7)           &1.4 (1.5)              &6.4 (6.5)         \\
        2.4         &0.20           &3600           &0.3        &$1^{\rm st}$ $r$-peak                                                          &15         &0.68 (11)           &--                     &3.7 (3.8)         \\
        \addlinespace[1ex]
        {\bf 3.4}   &{\bf 0.15}     &{\bf 3400}     &{\bf 4.0}  &{$\boldsymbol{Y_{\rm e} - 0.29}$\textbf{a} ($\boldsymbol{1 \leq Z \leq 92}$)}  &{\bf 300}  &{\bf 6.1 (9.9)}  &{\bf 2.4 (2.4)}  &{\bf 22 (22)}    \\
        3.4         &0.15           &3400           &4.0        &\AngIII\ ($1 \leq Z \leq 56$)                                                  &300        &6.1 (9.9)        &2.4 (2.4)        &22 (22)          \\
        3.4         &0.15           &3400           &0.5        &\AngII\                                                                        &38         &1.8 (7.4)        &1.1 (1.2)        &4.9 (4.9)        \\
        3.4         &0.15           &3400           &0.3        &$1^{\rm st}$ $r$-peak                                                          &23         &3.9 (17)         &--               &5.7 (5.7)        \\
        \addlinespace[1ex]
        {\bf 4.4}   &{\bf 0.12}     &{\bf 3200}     &{\bf 4.0}  &{$\boldsymbol{Y_{\rm e} - 0.29}$\textbf{a} ($\boldsymbol{1 \leq Z \leq 92}$)}  &{\bf 380}  &{\bf 11 (12)}  &{\bf 3.0 (3.1)}  &{\bf 28 (28)}  \\
        4.4         &0.12           &3200           &4.0        &\AngIII\ ($1 \leq Z \leq 56$)                                                  &380        &11 (12)        &3.0 (3.1)        &28 (28)        \\
        4.4         &0.12           &3200           &0.5        &\AngII\                                                                        &48         &5.9 (9.3)      &1.5 (1.5)        &6.2 (6.2)      \\
        4.4         &0.12           &3200           &0.3        &$1^{\rm st}$ $r$-peak                                                          &29         &13 (22)        &--               &7.2 (7.2)      \\
        \hline
    \end{tabular}
    \label{tab:TARDIS model parameters}
\end{table*}

\par
Despite achieving quite good agreement between the model and observations between $0.5 - 1.25$\,\micron, and in the region of the \SrII\ absorption feature, there are noticeable discrepancies between the model and the data. Recall that the spectrum is accurately calibrated to photometric measurements across the full wavelength range. The disagreement in the NIR is likely due to the photospheric approximation within \tardis\ (see Section~\ref{sec:Spectral analysis method}.) In our \tardis\ models, we have tuned the position of the photosphere such that it agrees with the data in the UV and optical parts of the spectrum. The code does not simulate anything beneath this boundary, and so we lose any information from NIR photons beneath this boundary. Therefore, we expect our model continua at wavelengths, \mbox{$\lambda \gtrsim 12000$\,\AA}, to be under-luminous relative to observations, since our NIR photon count is lower than we expect from a real astrophysical explosion. Because of this, we highlight that any difference between observations and our \tardis\ models beyond this wavelength should not be interpreted as support for a real NIR excess in the data, in the sense of the excess proposed by two-component kilonova models \citep[as in][]{Chornock2017, Cowperthwaite2017, Kasen2017, Coughlin2018}.

\par
The disagreement in the UV and optical part of the spectrum is more informative with respect to the composition of the ejecta. Our \AngII\ model spectrum exhibits too much absorption below $\sim 5000$\,\AA. We have tuned the amount of material in the simulation (through the density parameter $\rho_0$) such that the \SrII\ feature matches the spectrum shape, but this leads to too much material causing significant absorption in the blue, which is not observed. Assuming that the feature between \mbox{$\sim 7000 - 10000$\,\AA} really is \SrII, this implies that the amount of Sr in our composition is too low, relative to the other elements in the \AngII\ composition profile. If we drop the total amount of material in the ejecta to reduce absorption in the blue (by reducing $\rho_0$), and better match the data at wavelengths $\lesssim 6000$\,\AA, the \SrII\ absorption feature becomes much too weak. This strongly implies that the data require a composition with higher Sr abundance than that in \AngII\ (e.g. \AngXI\ and \AngXII).

\subsubsection{Other \Ye\ compositions} \label{sec:Spectral analysis results - Epoch 1 - Other Ye compositions}
\par
A similarly good fit can be obtained using the \AngXI\ composition, with parameters similar to those used for the best-fitting \AngII\ model presented in Figure~\ref{fig:1.4d Ang02 model} (\mbox{$v_{\rm min} = 0.28\,c$}, \mbox{$T = 4500$\,K}, and \mbox{$\rho_{0} = 10^{-14}$\gcm} for the \AngXI\ model, versus \mbox{$v_{\rm min} = 0.28\,c$}, \mbox{$T = 4500$\,K}, and \mbox{$\rho_{0} = 1.2 \times 10^{-14}$\gcm} for our best-fitting \AngII\ model). This is to be expected, since \AngII\ and \AngXI\ exhibit similar compositions, at least across the dominant few elements (see Figure~\ref{fig:Mass fraction compositions plot} and Table~\ref{tab:Ye bins (top 10)} for a comparison between the ten most abundant elements in each of the \Ye\ composition profiles). The abundance of Sr increases by a factor $\sim 1.7$ from \AngII\ to \AngXI, which is somewhat reflected in the drop in $\rho_0$ needed to fit the \SrII\ feature in the observed spectrum with these composition profiles.

\par
\begin{figure}
    \centering
    \includegraphics[width=\linewidth]{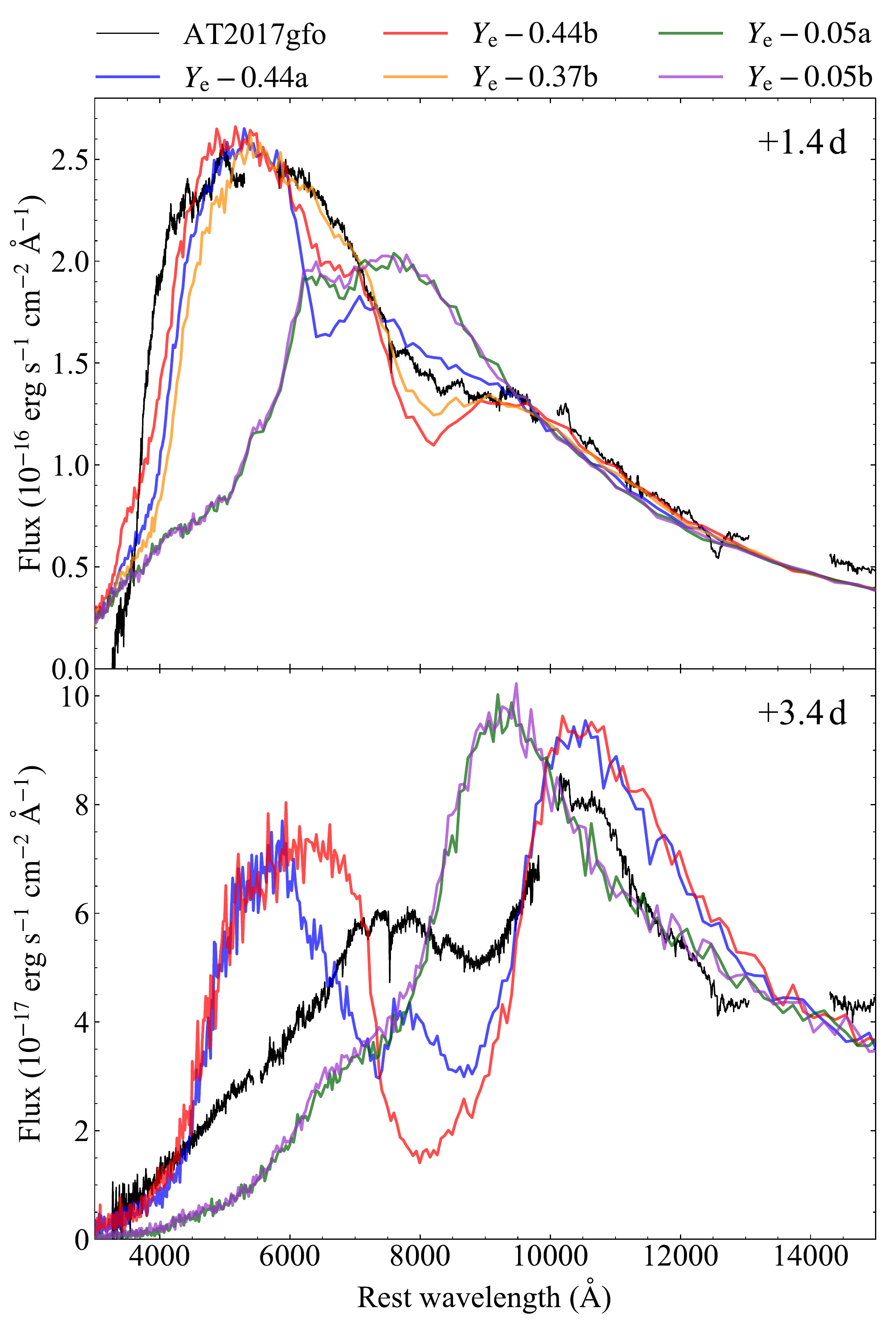}
    \caption{
        \xsh\ spectra of \gfo, taken 1.4 days (top panel) and 3.4 days (bottom panel) after the GW trigger, plotted alongside models with different \Ye\ compositions. The \AngI, \AngXII, \AngXI, \AngVI\ and \AngVII\ models are plotted in blue, red, orange, green and purple, respectively. \textit{Top panel:} +1.4\,d \xsh\ spectrum of \gfo\ plotted alongside different composition models. The \AngI, \AngXII, \AngVI\ and \AngVII\ models retained model parameters identical to our best-fitting \AngII\ model presented in Figure~\ref{fig:1.4d Ang02 model}, while our \AngXI\ model had a lower value for $\rho_0$ ($10^{-14}$\gcm), as discussed in the main text (see Section~\ref{sec:Spectral analysis results - Epoch 1 - Other Ye compositions}). \textit{Bottom panel:} +3.4\,d \gfo\ spectrum compared with different composition models. All models presented (\AngI, \AngXII, \AngVI\ and \AngVII) retain model parameters identical to the best-fitting \AngIII\ model, presented in Section~\ref{sec:Spectral analysis results - Epoch 3}.
    }
    \label{fig:1+3d Ang01+06+07+11+12 models}
\end{figure}

However, this \AngXI\ model suffers from the same issue as our best-fitting \AngII\ model, namely that there is still too much absorption in the blue, which can be mostly attributed to Y and Zr (see the SDEC plot in the lower panel of Figure~\ref{fig:1.4d Ang02 model}). While the increase in Sr abundance from \AngII\ to \AngXI\ is favourable (0.194 to 0.331), the accompanying abundance increase in Zr (0.129 to 0.166) is disfavourable. The Y abundance decreases slightly, from 0.031 in \AngII, to 0.029 in \AngXI. This small decrease does not significantly impact our model spectra. Although the increase in Sr abundance means we can now fit the \SrII\ absorption feature with less total ejecta material, an increase in the abundance of Zr counteracts any intended improvement we would hope to see in the blue end of the spectrum. We plot this \AngXI\ model alongside the +1.4\,d \xsh\ spectrum in the upper panel of Figure~\ref{fig:1+3d Ang01+06+07+11+12 models}.

\par
This pronounced line-blanketing effect seen in our models is not too surprising. Both Y and Zr are open $d$-shell elements, with many low-lying energy levels, which in turn leads to significant absorption. \cite{Kawaguchi2021} show that the opacity contribution from Y and Zr (as well as the lanthanides), can act as strong flux suppressors at near-UV and optical wavelengths. \cite{Ristic2022} conclude the same, at least for the Zr case.

\par
For completeness, we explore what our models predict with our high (\AngI\ \& \AngXII) and low (\AngVI\ \& \AngVII) \Ye\ composition profiles. We plot the model spectra corresponding to our best-fitting model parameters (for our best-fitting \AngII\ model presented in Section~\ref{sec:Spectral analysis results - Epoch 1 - Best-fitting Ye composition}), with both \AngI\ and \AngXII\ composition profiles, in the upper panel of Figure~\ref{fig:1+3d Ang01+06+07+11+12 models}, alongside the +1.4\,d \xsh\ spectrum of \gfo. The composition profiles represented by \AngI\ and \AngXII\ have the lightest element compositions, and are relatively rich in Fe-group elements. These model spectra both produce an absorption feature centred at \mbox{$\sim 6500$\,\AA}. This is produced by the \mbox{Ca\II\ $3\rm{d} - 4\rm{p}$} NIR triplet. This Ca\II\ triplet is analogous to the \SrII\ triplet due to their similar electronic configurations, and is commonly observed in supernova spectra. Despite Ca only being present in small quantities in our \AngI\ and \AngXII\ compositions ($\sim 1.5$ and 0.4 per cent, respectively), the triplet transitions are prominent in the model spectrum, and act as a strong argument against the high \Ye\ composition profiles. This result is in agreement with that presented by \cite{Domoto2021}, where they found that the \CaII\ NIR triplet was also prominent in their lanthanide-poor KN models. Although these models are a poor fit to the data, they do serve to highlight what future KN events may look like at this epoch (assuming it is blue and dominated by `light' ejecta material).

\par
We also calculated \tardis\ model spectra for the \AngVI\ and \AngVII\ composition profiles (which have \Xlanth\ = 0.38 and 0.34, respectively), and these models are plotted in the upper panel of Figure~\ref{fig:1+3d Ang01+06+07+11+12 models}. These models are significantly redder, as they have much stronger absorption in the optical, and they fail to produce the $\sim 1$\,\micron\ \SrII\ feature. Clearly these models do not match observation, but they do demonstrate how early spectra may appear if lanthanide-rich kilonovae are found in the future (and there is no blue component). These peak at wavelengths between \mbox{$\sim 7000 - 8000$\,\AA}, indicating the $i$-band is likely to be the `sweet spot' for searching for distant, lanthanide-rich kilonovae. Finally, we note that the heavy element \AngVI\ and \AngVII\ composition profiles contain significant platinum and gold, and, in agreement with \cite{Gillanders2021}, we find no signatures of these in the \gfo\ spectra.

\subsubsection{First \rpro\ peak model} \label{sec:Spectral analysis results - Epoch 1 - First r-process peak model}
\par
The dominant elements in our best-fitting \AngII\ model spectrum at the blue end ($\lambda \lesssim 6000$\,\AA) of the spectrum are Y and Zr. This is clear from looking at the SDEC plot in the lower panel of Figure~\ref{fig:1.4d Ang02 model}. Our model clearly exhibits strong absorption from Y and Zr at the blue end, and prominent Sr absorption between \mbox{$\sim 7000 - 10000$\,\AA}.

\par
From various \rpro\ studies \citep[see][for a review on the \rpro]{Arnould2007}, we expect that if Sr is synthesised, then these other two first \rpro\ peak elements must also be synthesised. However, if we tune the amount of Sr to match the data, while keeping the mass fractions fixed, then this leads to too much absorption from Y and Zr. This implies that the relative ratios of \mbox{Sr : Y : Zr} in our \AngII\ composition (and, by extension, in \AngXI) may not represent what is observed in \gfo. We thus generated models consisting purely of these three species, and freely varied their relative ratios, and $\rho_0$, while keeping all other model parameters from our best-fitting \AngII\ model fixed.

\par
We found that to fit the observations, we need a value of \mbox{$\rho_0 = 4 \times 10^{-15}$\gcm}, a factor of 3 lower than our best-fitting \AngII\ model. We also found that we need more Sr than both Y and Zr combined. We favour a simple model consisting of 75 per cent Sr, no Y, and 25 per cent Zr (a ratio of \mbox{$3 : 0 : 1$}), whereas, for comparison, \AngII\ contains 19.4 per cent Sr, 3.1 per cent Y, and 12.9 per cent Zr (\mbox{$\sim 6 : 1 : 4$}), and \AngXI\ contains 33.1 per cent Sr, 2.9 per cent Y, and 16.6 per cent Zr (\mbox{$\sim 11 : 1 : 6$}). The solar \rpro\ composition (calculated by taking the total solar abundance from \citealt{Asplund2009}, and subtracting off the $s$-process contribution from \citealt{Bisterzo2014}) has relative ratios of \mbox{$\sim 6 : 1 : 3$}. We note that there is some degeneracy between Y and Zr in our models, since they both contribute to the blanket absorption at the blue end of the spectrum. As such, with this new best-fitting model composition, we emphasise that we are \textit{not} claiming that there is zero Y present in the ejecta of \gfo; we simply find that we can adequately reproduce the overall shape of the spectrum with just Sr and Zr (see the discussion in Section~\ref{sec:Discussion and interpretation - Ionisation and atomic data uncertainties} on the uncertainty in \SrII\ ionisation). 

\par
Figure~\ref{fig:1.4d Ang02 model} compares the best-fitting \AngII\ \tardis\ model with the +1.4\,d \xsh\ spectrum of \gfo. Plotted alongside is our best-fitting model composed purely of Sr and Zr, which we refer to as our `$1^{\rm st}$ $r$-peak' model. Removal of additional species outside of the first \rpro\ peak, as well as adjusting the relative abundances of the first \rpro\ peak elements, improves the agreement between the model and observations. Quite a satisfactory fit is found with this simple composition.

\subsection{Epoch 2: +2.4 day spectrum} \label{sec:Spectral analysis results - Epoch 2}
\par
\begin{figure}
    \centering
    \includegraphics[width=\linewidth]{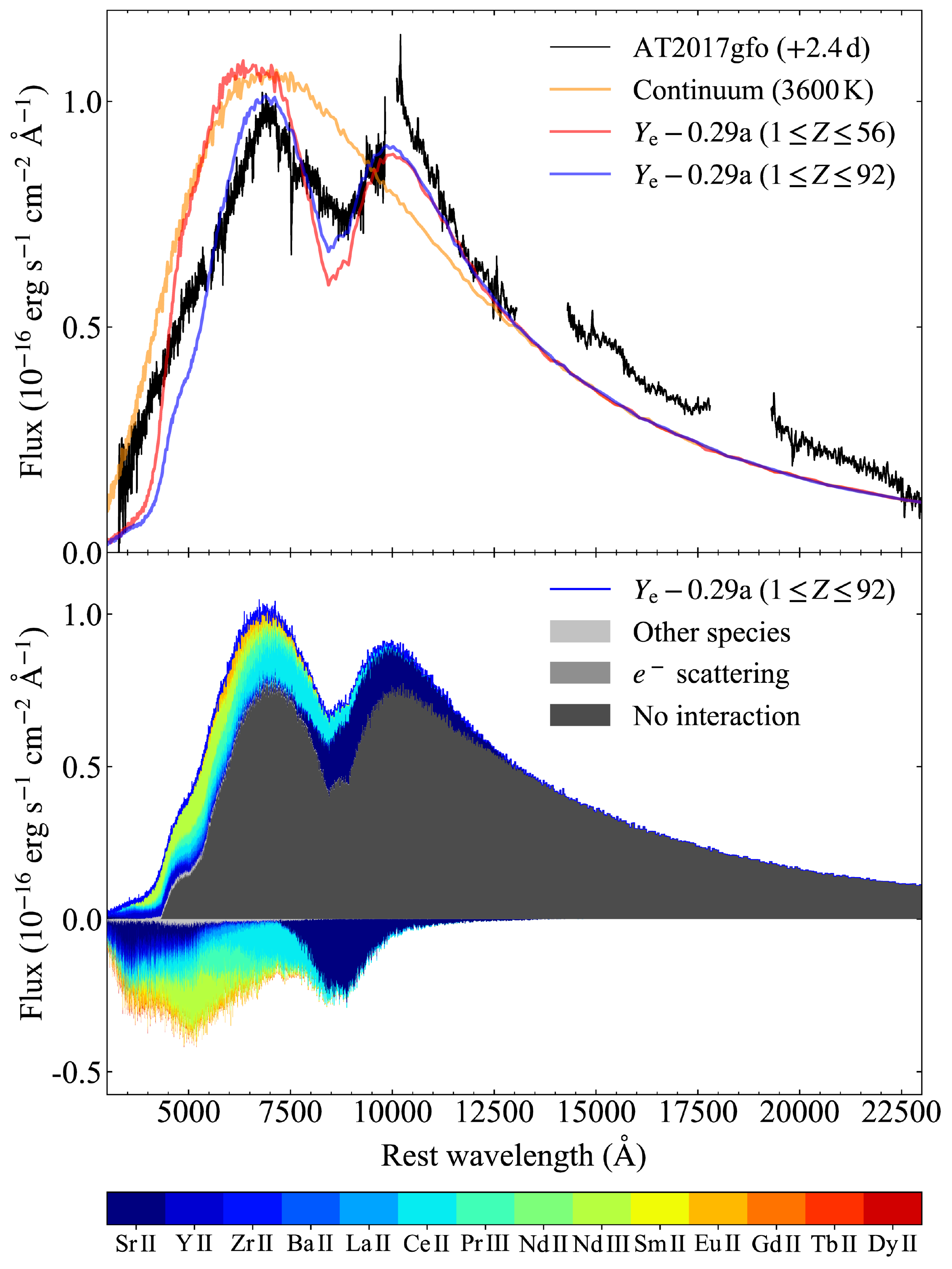}
    \caption{
        \xsh\ spectrum of \gfo, taken 2.4\,d after explosion, plotted alongside the best-fitting \AngIII\ model (blue), the continuum from this best-fitting model (orange), and the best-fitting \AngIII\ model, but with all transitions originating from species with $Z > 56$ removed (red).
    }
    \label{fig:2.4d Ang03 model}
\end{figure}

The second \xsh\ spectrum of \gfo, taken +2.4\,d after the GW trigger, exhibits a significantly redder SED than that observed a day earlier. This rapid colour evolution is most pronounced between the first and second epochs; subsequent spectra have SEDs similar to this +2.4\,d spectrum. Here, we can broadly reproduce the SED with a much cooler continuum, where \mbox{$T = 3600$\,K}. Our best-fitting composition profile for this epoch is \AngIII. This composition profile contains significantly more heavy elements than \AngII\ (a factor of $\sim 430$ higher), which produce strong line-blanketing at the bluer wavelengths. It is this strong absorption that produces the observed flux suppression at wavelengths $\lesssim 7000$\,\AA. Figure~\ref{fig:2.4d Ang03 model} shows our best-fitting \AngIII\ model compared to observation. We are again able to reproduce the strong absorption feature between \mbox{$\sim 7000 - 10000$}\,\AA\ with the \SrII\ triplet. Fitting this feature with Sr constrains the minimum ejecta velocity, \mbox{$v_{\rm{min}} = 0.20\,c$}, and \mbox{$\rho_0 = 4.0 \times 10^{-15}$\gcm}. This corresponds to a \SrII\ mass, \mbox{\MSrII\ $= 1.9 \times 10^{-6}$}\,\msun.

\par
The \tardis\ model produces a prominent P-Cygni line profile from the \SrII\ transitions at this epoch, with the model emission component peaking at $\sim 1.0$\,\micron. It provides a reasonable but not perfect match to the shape of the observed spectrum, and indicates that even a small amount of Sr results in a strong observed feature. Recall that we are using the \texttt{macroatom} line interaction treatment, which accounts for fluorescence effects that may impact the relative strengths of the absorption and emission components of this \SrII\ P-Cygni feature. This further supports the results presented by \cite{Watson2019}, who proposed \SrII\ as the ion producing the absorption and the P-Cygni line in these first two \xsh\ spectra, although their \tardis\ models did not produce a strong emission component. Our match to the continuum in the NIR suffers from the same issue as discussed previously in Sections~\ref{sec:Spectral analysis method} and \ref{sec:Spectral analysis results - Epoch 1 - Best-fitting Ye composition}, and cannot necessarily be interpreted as a NIR excess.

\par
As in Section~\ref{sec:Spectral analysis results - Epoch 1 - Best-fitting Ye composition}, we also plot the continuum from our best-fitting model to highlight the effects of the line-forming region on the emergent spectrum. To demonstrate the important effect that the heavy \rpro\ elements have on our best-fitting model, and continuum suppression at blue wavelengths, we remove all transitions belonging to any species heavier than Ba (\mbox{$Z > 56$}), and re-generate our best-fitting model. Figure~\ref{fig:2.4d Ang03 model} shows that without the inclusion of the lanthanides \mbox{($57 \leq Z \leq 70$)}, we cannot reproduce the flux distribution below $\sim 7000$\,\AA\ in the observed spectrum. The ions that dominate the line blanketing are the lanthanides \CeII, \NdII, \NdIII, \SmII\ and \EuII, and the lanthanide mass fraction in this composition is \Xlanth~$= 0.05$. This result is in contrast to the \tardis\ models presented by \cite{Watson2019}, where they proposed that no heavy element ($Z > 38$) contribution was necessary in their models to reproduce the blue ends of the $+2.4 - 4.4$\,d \xsh\ spectra.

\subsection{Epoch 3: +3.4 day spectrum} \label{sec:Spectral analysis results - Epoch 3}
\par
The evolution of the SED of \gfo\ between epochs~2~and~3 is not as pronounced as that observed between epochs~1~and~2. The spectrum still contains the broad P-Cygni feature between \mbox{$\sim 7000 - 10000$\,\AA}. It also appears to have strong flux suppression at the blue end of the spectrum. We were again able to fit this epoch with the \AngIII\ composition, but with cooler ejecta, and a slower-moving inner boundary (\mbox{$T = 3400$\,K} and \mbox{$v_{\rm{min}} = 0.15\,c$}). We find a reasonable match to the feature between \mbox{$\sim 7000 - 10000$\,\AA} with the same $\rho_0$ value as the previous epoch (\mbox{$\rho_0 = 4.0 \times 10^{-15}$\gcm}), which corresponds to a \SrII\ mass, \mbox{\MSrII\ $= 6.1 \times 10^{-6}$}\,\msun\ in the line-forming region. 

\par
\begin{figure}
    \centering
    \includegraphics[width=\linewidth]{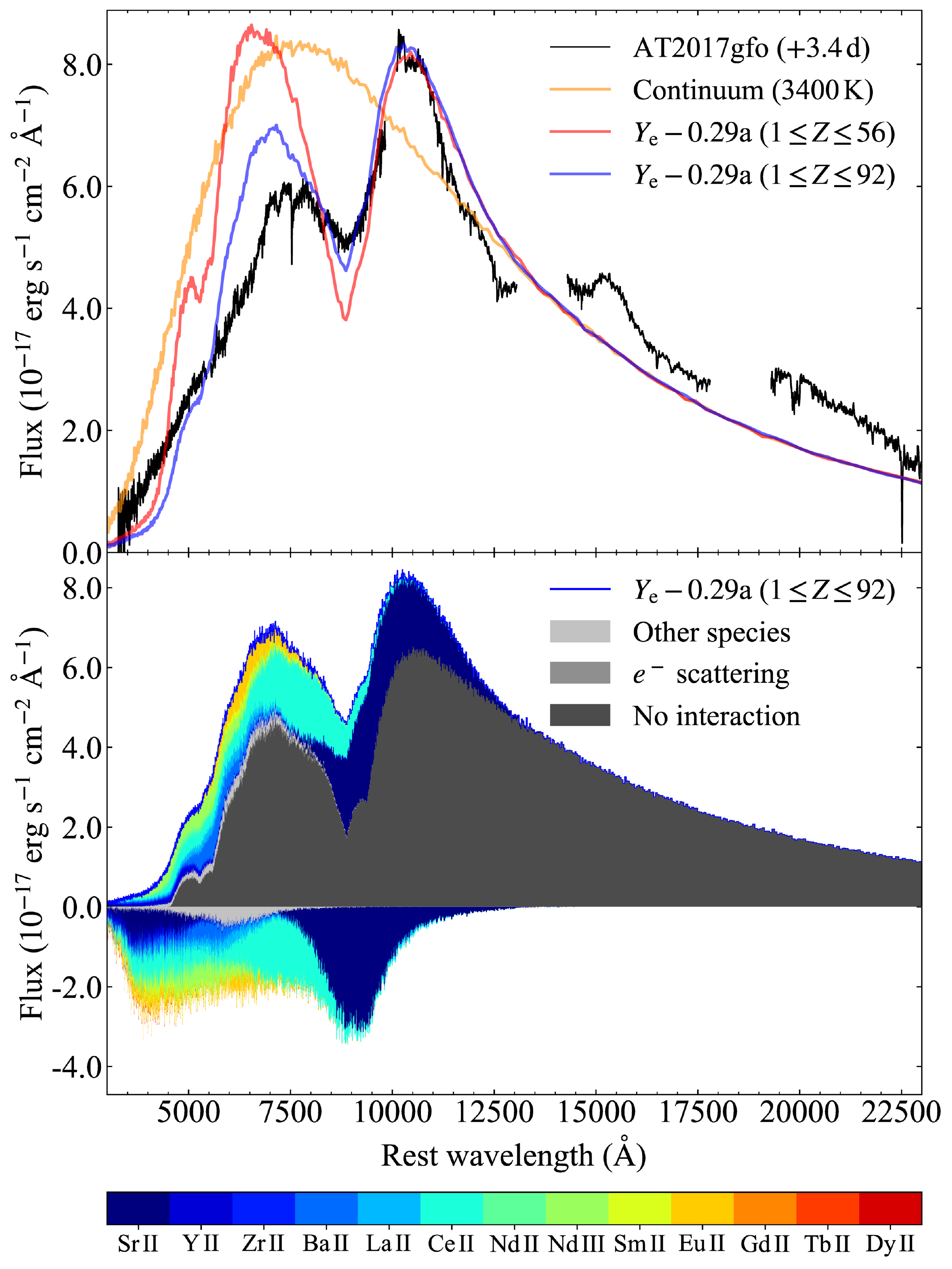}
    \caption{
        \xsh\ spectrum of \gfo, obtained 3.4\,d after explosion, plotted alongside the best-fitting \AngIII\ model (blue), the model continuum (orange), and the same model but with all transitions from species with $Z > 56$ removed (red). 
    }
    \label{fig:3.4d Ang03 model}
\end{figure}

The shape of the spectrum at wavelengths $\lesssim 6000$\,\AA\ is not matched exactly, but the lanthanides in the \AngIII\ composition play a significant role. The best-fitting model is compared to the observed spectrum in Figure~\ref{fig:3.4d Ang03 model}. As in Section~\ref{sec:Spectral analysis results - Epoch 2}, we generate a comparison model with no lines from species with $Z > 56$, to highlight the importance of these heavy species. We propose that the presence of lanthanides is a requirement to suppress the flux between $\sim 4500 - 7500$\,\AA\ through line blanketing. Without these ions the spectrum is significantly different, with elevated flux levels in the blue, as demonstrated in Figure~\ref{fig:3.4d Ang03 model}. The NIR region of the spectrum suffers from the same issue as the previous models, and we note that there are now two emission features prominent above the continuum, to which we will return in \paperII.

\par
As in Section~\ref{sec:Spectral analysis results - Epoch 1 - Other Ye compositions}, we generate some models for our high (\AngI\ \& \AngXII) and low (\AngVI\ \& \AngVII) \Ye\ composition profiles, to demonstrate that neither are viable compositions (see the lower panel of Figure~\ref{fig:1+3d Ang01+06+07+11+12 models}). As in the +1.4\,d case, the \AngI\ model produces a strong absorption feature due to the \CaII\ NIR triplet. This feature is not prominent in the \AngXII\ model, likely because of the lower Ca abundance ($\sim 1.5$ per cent in \AngI, versus $\sim 0.4$ per cent in \AngXII). Both of the high \Ye\ compositions exhibit a very pronounced \SrII\ NIR triplet absorption feature, which is much stronger than the equivalent feature in the observed spectrum at this epoch. Our low \Ye\ models (\AngVI\ and \AngVII) have much stronger absorption in the optical, and fail to produce the $\sim 1$\,\micron\ emission feature (likely the \SrII\ triplet P-Cygni emission). This is due to the absence of any substantial amount of Sr material in these composition profiles. They have a strong flux deficit in the blue and a peak at $\sim 9000$\,\AA\ (in the $z-$band filter). This set of models act to highlight what future observations of either lanthanide-poor or lanthanide-rich KN events may look like, a few days after explosion.

\subsection{Epoch 4: +4.4 day spectrum} \label{sec:Spectral analysis results - Epoch 4}
\par
\begin{figure}
    \centering
    \includegraphics[width=\linewidth]{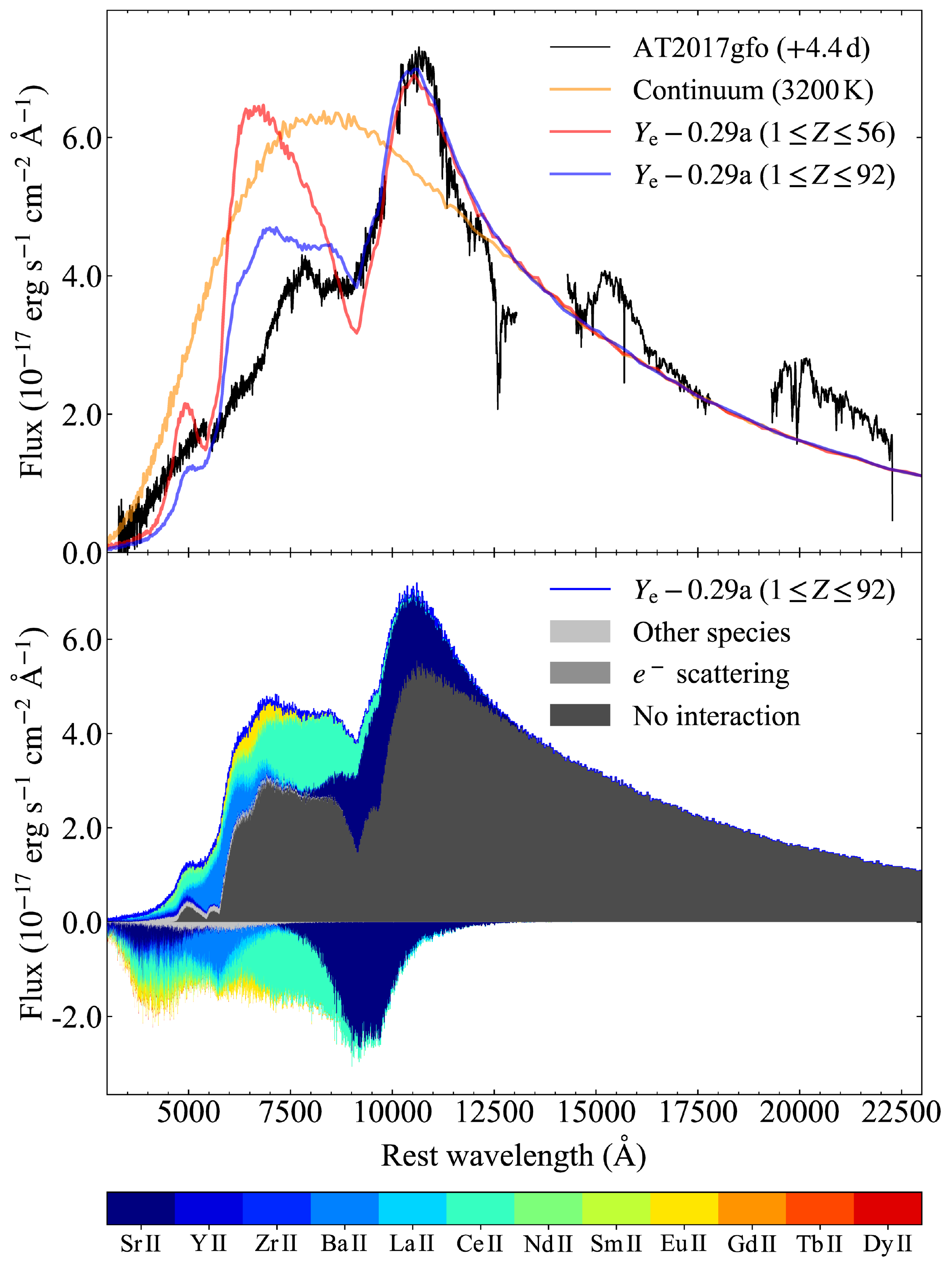}
    \caption{
        \xsh\ spectrum of \gfo, obtained 4.4\,d post-explosion, plotted alongside the best-fitting \AngIII\ model (blue), the model continuum (orange), and the same model but with all transitions from species with $Z > 56$ removed (red).
    }
    \label{fig:4.4d Ang03 model}
\end{figure}

Due to the photospheric approximation within \tardis, the code is incapable of reliably modelling the spectra of \gfo\ beyond the time when the inner regions of the ejecta become optically thin. The rapid evolution of \gfo\ could potentially present a problem with interpreting the spectra beyond a few days. It is not immediately clear beyond what epoch the spectra can no longer be adequately modelled with a single-temperature blackbody photosphere. Hence, for the +4.4\,d spectrum, we do not focus on iterating over model parameters in detail, but simply evolve our best-fitting model from the second and third epoch. We modified only the ejecta temperature and minimum velocity of the ejecta to which we are sensitive, and find reasonable agreement between the observed spectrum and our \tardis\ model, using \mbox{$T = 3200$\,K} and \mbox{$v_{\rm{min}} = 0.12\,c$} (and using the \AngIII\ composition with \mbox{$\rho_0 = 4.0 \times 10^{-15}$\gcm}). The resultant best-fitting model is presented in Figure~\ref{fig:4.4d Ang03 model}.

\par
The strong absorption feature, present in the $+1.4 - 3.4$\,d spectra between \mbox{$\sim 7000 - 10000$\,\AA} has evolved into a strong and prominent P-Cygni feature that is in net emission. The lower panel of Figure~\ref{fig:4.4d Ang03 model} contains the SDEC plot for our best-fitting \AngIII\ model, and this P-Cygni feature is clearly still reproduced in our model spectra by the \SrII\ triplet, as in the previous epochs. With a \SrII\ mass, \mbox{\MSrII\ $= 1.1 \times 10^{-5}$\,\msun}, our model seems to obtain reasonable agreement with the emission component, with an associated weak absorption component. The spectrum shape around the absorption trough is not quantitatively well reproduced by our model, although the velocity distribution of the material broadly appears to represent what is observed. The shape of the line-blanketed region of the spectrum at wavelengths $\lesssim 7500$\,\AA\ does not exactly match the observed data, but again, the inclusion of a lanthanide mass fraction, \Xlanth~$= 0.05$, in the \AngIII\ composition is required to significantly suppress the otherwise blue flux that would be produced by a continuum with $T = 3200$\,K.

\par
As we have done for the previous two epochs, in Figure~\ref{fig:4.4d Ang03 model} we plot the best-fitting \AngIII\ model, but with all contributions belonging to elements with $Z > 56$ removed, illustrating the essential requirement of strong line absorption by these heavy elements, and the lanthanides in particular. Given the uncertainties in the line lists and atomic data for the lanthanides, their heavy influence on the data, and uncertainty of the photospheric assumption at this epoch, we suggest that this is a satisfactory reproduction of the data. The NIR region of the spectrum suffers from the same issue as previous models, and the two emission features noted in the previous epoch spectrum are still prominent above the continuum. We discuss these further in \paperII.

\subsection{Epochs 5 -- 7: +5.4 -- 7.4 day spectra} \label{sec:Spectral analysis results - Epochs 5-7}
\par
In Section~\ref{sec:Spectral analysis results - Epoch 4}, we highlighted the uncertainty that potentially arises from the photospheric approximation within \tardis. It is not clear whether beyond $\sim 3$ days \gfo\ remains in a fully photospheric regime \citep[also see discussion in][]{Gillanders2021}. Despite this, we managed to find a reasonable match to the spectrum at +4.4 days (Figure~\ref{fig:4.4d Ang03 model}), which we take to imply that the material is still in the diffusion phase. In a similar vein, we evolve our best-fitting \tardis\ model further in time, to explore how well our models continue to resemble the observed spectra. To this end, we forward-evolve our \tardis\ models to about one week post-explosion.

\par
\begin{figure}
    \centering
    \includegraphics[width=\linewidth]{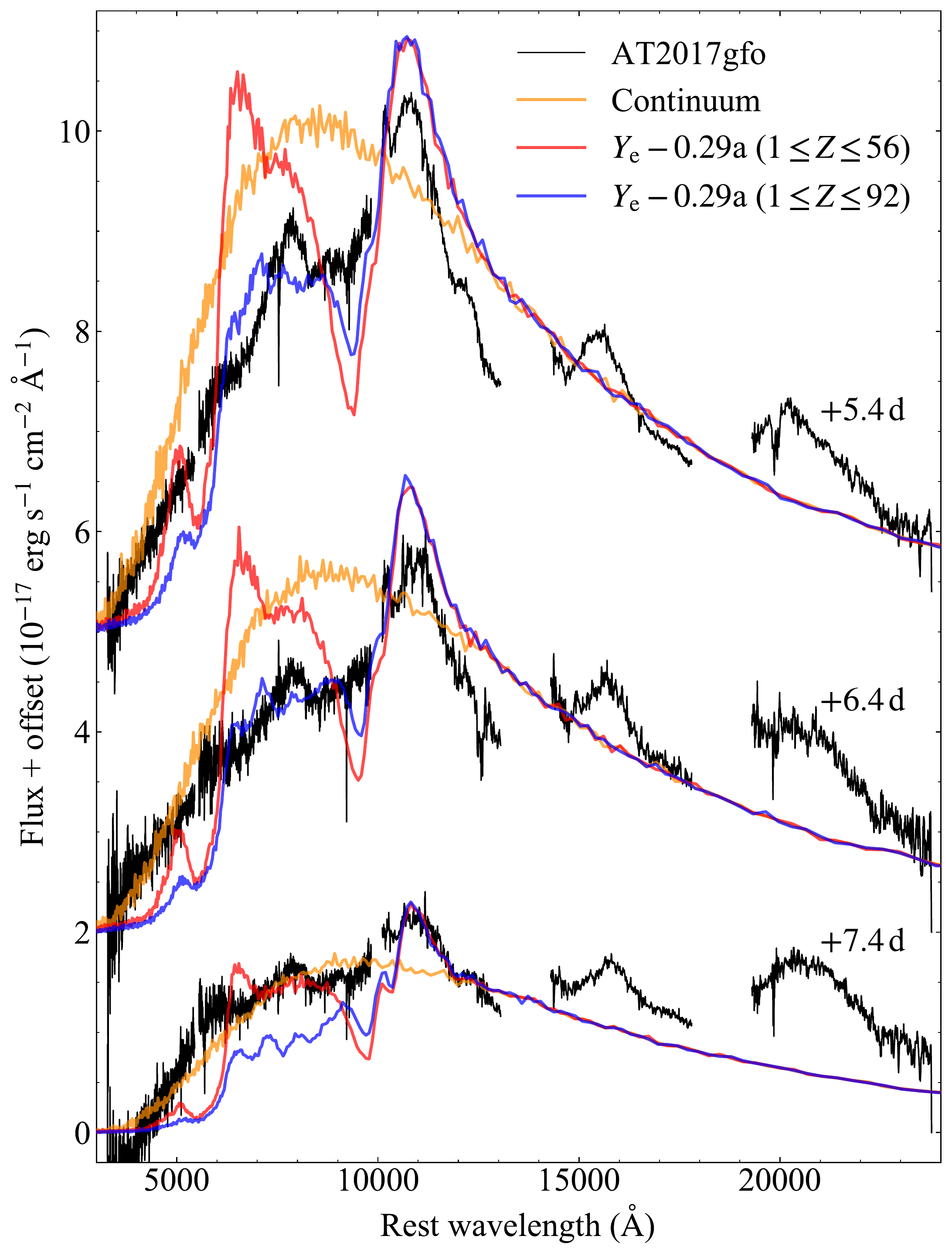}
    \caption{
        +5.4, +6.4 and +7.4\,d \xsh\ spectra of \gfo, plotted alongside their best-fitting \AngIII\ models (blue), their model continua (orange), and the same models but with all transitions from species with $Z > 56$ removed (red). The +5.4, +6.4 and +7.4\,d models have temperatures, \mbox{$T$ = 3200, 3100 and 2900\,K}, and inner boundary velocities, \mbox{$v_{\rm{min}}$ = 0.09, 0.07 and 0.05\,$c$}, respectively. All three models have the same value for $\rho_0$ ($4 \times 10^{-15}$\gcm). The observed spectra and corresponding models have been offset for clarity (\mbox{$5 \times 10^{-17}$\,\ergscmA} for +5.4\,d, and \mbox{$2 \times 10^{-17}$\,\ergscmA} for +6.4\,d).
    }
    \label{fig:5-7d Ang03 models}
\end{figure}

Our best-fitting models for the \xsh\ spectra of \gfo\ at +5.4, +6.4 and +7.4\,d are shown in Figure~\ref{fig:5-7d Ang03 models}. Our values for $v_{\rm{min}}$ continue to decrease, as is the trend exhibited across all epochs we have modelled thus far. The temperature also follows a general trend of decreasing with time, apart from the fifth epoch, where we find good agreement with the same temperature as the previous epoch ($T = 3200$\,K, $v_{\rm{min}} = 0.09\,c$ and \mbox{$\rho_0 = 4 \times 10^{-15}$\gcm}). As in the previous cases, we also plot the continua of our best-fitting models, and our models with all transitions from species with $Z > 56$ removed. These model spectra are all plotted in Figure~\ref{fig:5-7d Ang03 models}, alongside the +5.4\,d \gfo\ spectrum, for comparison.

\par
Comparing with the +5.4\,d spectrum of \gfo, the model spectrum still produces a strong \SrII\ triplet feature. This is in reasonable agreement with the persistent feature at the same wavelength in the observational data. However, the absorption in the \tardis\ model is much stronger than that in the observed spectrum (if indeed the observed spectrum even exhibits absorption). Clearly there is either less absorption in the observations than in our model, or there is something `boosting' the flux in the observed data at the position of the \SrII\ triplet absorption. As has been shown for the previous epochs, removing the heavy elements ($Z > 56$) negatively impacts the fit to the data at wavelengths $\lesssim 8000$\,\AA.

\par
The evolution through to +6.4\,d follows the same trend as previous epochs, with a good fit obtained using the \AngIII\ composition profile, and the same density profile as before. The values for $T$ and $v_{\rm min}$ decrease, as expected. Our best-fitting model (\mbox{$T = 3100$\,K}, \mbox{$v_{\rm{min}} = 0.07\,c$} and \mbox{$\rho_0 = 4 \times 10^{-15}$\gcm}) is presented in Figure~\ref{fig:5-7d Ang03 models}. We also illustrate a model with all elements with $Z > 56$ removed from the line-forming region. As previously demonstrated, there is significantly more flux produced at wavelengths \mbox{$\sim 6000 - 8000$\,\AA} when these heavy elements (specifically the lanthanides) are removed. While we are unsure if the inner boundary approximation within our \tardis\ model is still valid, this model broadly resembles the observed spectrum. However, it is beginning to fail to accurately reproduce the shape around the feature we attribute to \SrII\ at earlier epochs. The absorption present in the model P-Cygni \SrII\ feature is not present at all in the observed spectrum. Additionally, the emission component is too strong, compared with the observational data.

\par
Finally, we present our model evolved forward to the same epoch as the +7.4\,d \xsh\ spectrum of \gfo. At this epoch, the agreement between our model and the observed spectrum is poor. Our model produces a \SrII\ emission feature that peaks at the same wavelength as the prominent emission feature in the observed data, but the shape of the blue side of the feature does not match observation. Additionally, the model still shows an absorption component from the \SrII\ triplet, while there is no longer any evidence of absorption in this region of the observed spectrum. For this model, we have \mbox{$T = 2900$\,K}, \mbox{$v_{\rm{min}} = 0.05\,c$} and \mbox{$\rho_0 = 4 \times 10^{-15}$\gcm}. In general, beyond matching the peak of the \SrII\ feature, the model does not agree with the observations.

\par
The models appear to no longer capture the broad appearance of \gfo\ beyond $\sim 6 - 7$ days, and so we opted to cease modelling any subsequent spectra using \tardis. It appears that the spectra taken from $+0.5 - 4.4$\,d are able to be physically reproduced by \tardis, with its inner boundary, single-temperature blackbody approximation. The +5.4 and 6.4\,d spectra appear to be in a transition regime, where the code can still adequately reproduce the observations, with some caveats as highlighted above. However, by +7.4\,d, our \tardis\ model fit is poor, implying that \tardis\ is no longer suitable for modelling the ejecta. 

\par
The $H$ and $K$-band regions of the spectrum have developed two interesting features (as previously noted in Sections~\ref{sec:Spectral analysis results - Epoch 3}~and~\ref{sec:Spectral analysis results - Epoch 4}). There are two broad emission line profiles, centred in each band, which we will return to quantitatively in \paperII. Despite the difficulty \tardis\ has with reproducing a reliable continuum in the NIR (as discussed in Sections~\ref{sec:Spectral analysis method} and \ref{sec:Spectral analysis results - Epoch 1 - Best-fitting Ye composition}), these features appear to be broad line profiles in emission, enhancing the NIR flux over the pseudo-thermal continuum.

\section{Discussion \& Interpretation} \label{sec:Discussion and interpretation}
\subsection{Composition} \label{sec:Discussion and interpretation - Composition}
\par
Our analysis shows that the \xsh\ spectrum taken +1.4\,d after merger is notably different from the subsequent epochs. The observed spectrum is blue and the continuum from our two most closely fitting models (\mbox{$T = 4500$\,K} and \mbox{$v_{\rm min} = 0.28\,c$}) broadly reproduces the observed spectral shape from the near-UV through the optical region (Figure~\ref{fig:1.4d Ang02 model}). We can produce a fairly satisfactory model with a composition composed entirely of first \rpro\ peak species. We confirm that the broad absorption feature between $\sim 7000 - 10000$\,\AA\ can be reproduced by the \SrII\ NIR triplet, as first proposed by \cite{Watson2019}. However, our best-fitting model requires a ratio of first \rpro\ peak elements that deviates from the solar \rpro\ composition, with mass fractions of 75 per cent Sr, 25 per cent Zr and zero Y (which we labelled \mbox{`$1^{\rm st}$ $r$-peak'} in Section~\ref{sec:Spectral analysis results - Epoch 1 - First r-process peak model}). Employing the mass fractions in the \AngII\ model indicates that even modest amounts of Y (3.1 per cent) and Zr (12.9 per cent) are enough to cause excess absorption at wavelengths $\lesssim 5000$\,\AA, impairing the quality of the fit (Figure~\ref{fig:1.4d Ang02 model}). The ratio of these three elements is close to solar in our \AngII\ composition profile. Therefore, either the true composition is somewhat different to the solar ratio, or the ionisation fractions of these elements are not being captured correctly (which we explore further in Section~\ref{sec:Discussion and interpretation - Ionisation and atomic data uncertainties}).

\par
\begin{figure}
    \centering
    \includegraphics[width=\linewidth]{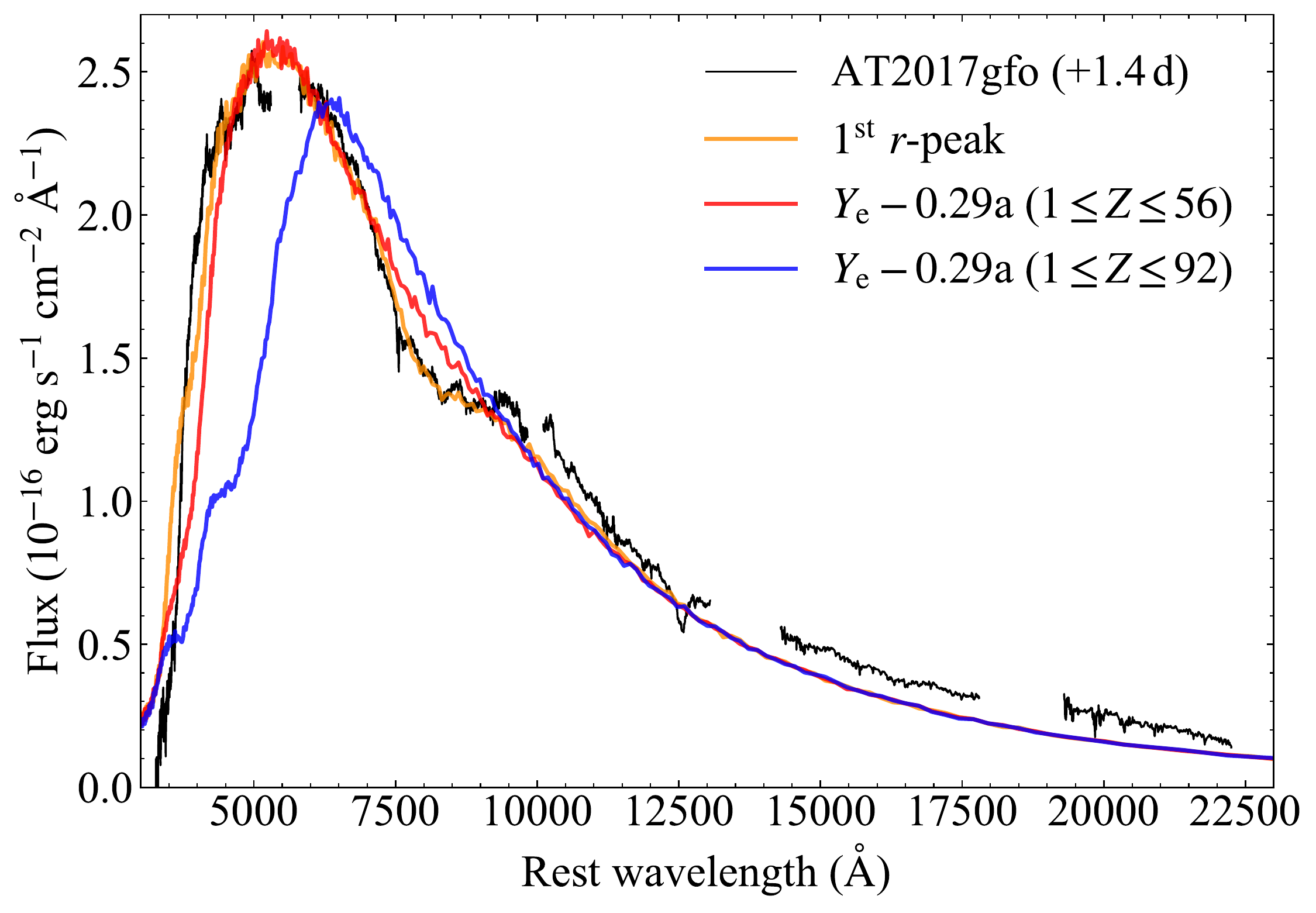}
    \caption{
        +1.4\,d \xsh\ spectrum of \gfo\ plotted alongside our \AngIII\ composition model (blue), and the same model, but with all transitions from species with $Z > 56$ removed (red). Also plotted is our best-fitting $1^{\rm st}$ $r$-peak model (orange), for comparison.
    }
    \label{fig:1.4d Ang03 model}
\end{figure}

The \AngII\ composition has a very low lanthanide mass fraction (\Xlanth\ $\simeq 10^{-4}$), and our $1^{\rm st}$ $r$-peak composition, by definition, has \Xlanth~$= 0$. The presence of even moderate quantities of heavy lanthanide species is enough to significantly alter the model at wavelengths $\lesssim 6000$\,\AA. To illustrate this, we generated a model with the \AngIII\ composition profile, which contains a modest lanthanide mass fraction (\Xlanth~$= 0.05$), and plot this against the +1.4\,d \xsh\ spectrum of \gfo\ (see Figure~\ref{fig:1.4d Ang03 model}). We also plot the same model, but with all transitions from species with $Z > 56$ removed, for completeness. Finally, we also include our best-fitting $1^{\rm st}$ $r$-peak \tardis\ model, presented in Section~\ref{sec:Spectral analysis results - Epoch 1 - First r-process peak model}. Clearly, even modest quantities of the complex lanthanide elements is enough to cause the model to deviate strongly from the observations. The absorption feature, which we attribute to the \SrII\ triplet feature, is not reproduced by this \AngIII\ model, as there is not enough \SrII\ present. Although increasing the total amount of material in the model (through increasing $\rho_{0}$) would produce this feature, the deviation between the model and observed spectra at wavelengths blueward of this \SrII\ feature would become even more pronounced.

\par
This indicates that there cannot be any significant quantity of lanthanide material present in the line-forming region that produces the +1.4\,d spectrum. Further analysis to constrain the amount of lanthanides that can be permitted at this epoch provided an upper limit on the mass fraction of lanthanide material, \mbox{\Xlanth\ $\lesssim 5 \times 10^{-3}$}. This is in reality expected to be lower, since, at this model temperature, most of our lanthanide material is doubly ionised, and our atomic data is quite sparse for these ions. All of the above supports the idea that the earliest optical emission (up to $\sim 1 - 2$ days after merger) is free from any significant quantity of lanthanide material, and is dominated by light (first \rpro\ peak) material.

\par
\begin{figure}
    \centering
    \includegraphics[width=\linewidth]{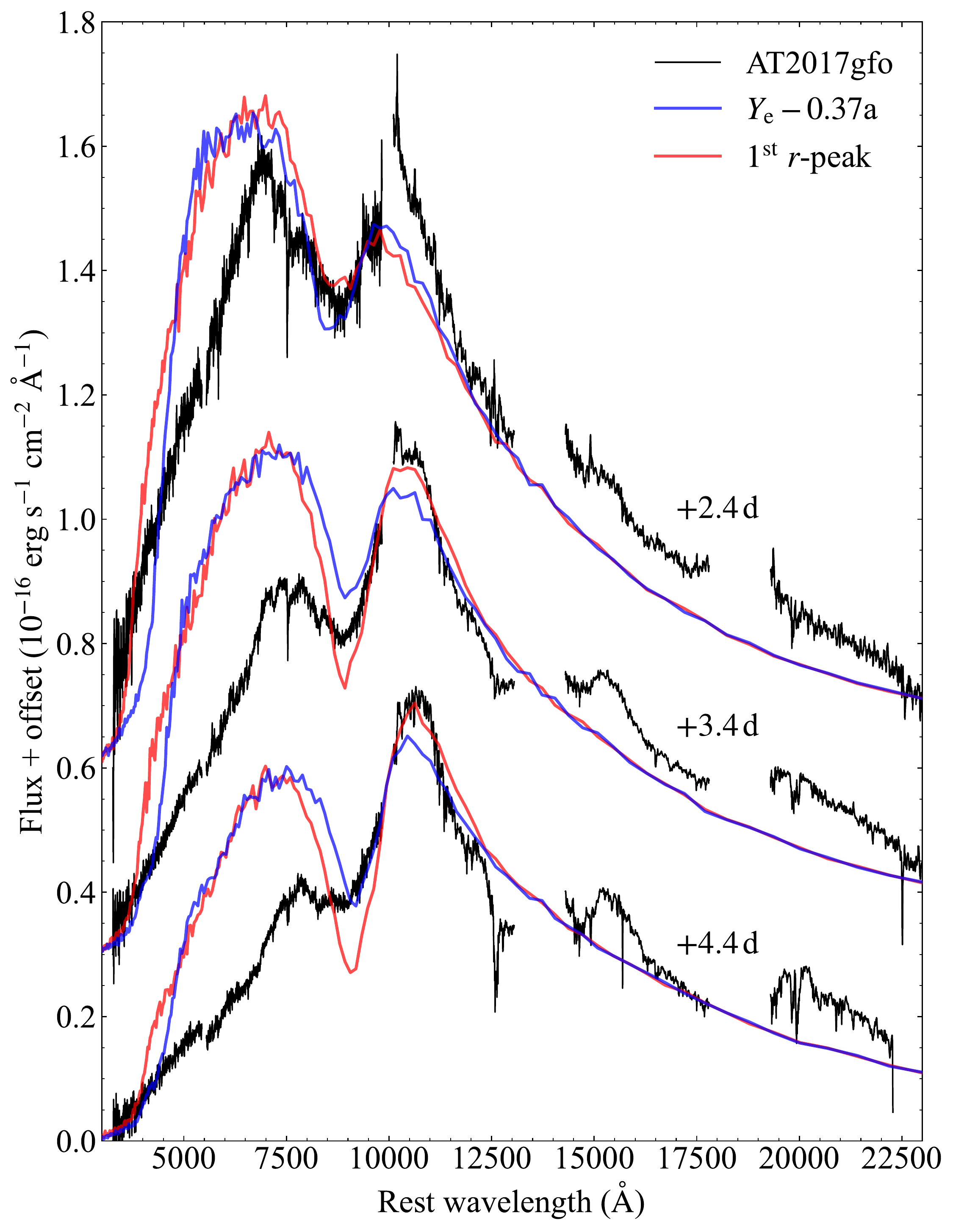}
    \caption{
        \xsh\ spectra of \gfo, taken +2.4, +3.4 and +4.4\,d after explosion, compared with our forward modelled best-fitting \AngII\ (blue) and $1^{\rm st}$ $r$-peak (red) models for the +1.4\,d \xsh\ spectrum. The observed spectra and corresponding models have been offset for clarity (\mbox{$0.6 \times 10^{-16}$\,\ergscmA} for +2.4\,d, and \mbox{$0.3 \times 10^{-16}$\,\ergscmA} for +3.4\,d).
    }
    \label{fig:2-4d Ang02 and Sr+Zr models}
\end{figure}

However, we find a distinct difference in the required composition for all subsequent epochs, compared to the first epoch composition. For the spectra at $+2.4 - 4.4$\,d, we find good agreement with the data using the \AngIII\ composition profile. We find that it is now \textit{necessary} to have some modest amount of lanthanide material (\Xlanth~$ = 0.05$) to produce the required level of line absorption at wavelengths $\lesssim 7500$\,\AA, to replicate the observed spectra. To highlight this point further, we generated models for epochs $2 - 4$, with the compositions that worked best for the first epoch spectrum (the \AngII\ composition profile with \mbox{\Xlanth\ $\simeq 10^{-4}$}, and our $1^{\rm st}$ $r$-peak model). Figure~\ref{fig:2-4d Ang02 and Sr+Zr models} shows that the models fail to match the data, and they suffer from the same issue as the \AngIII\ models with all transitions belonging to species with $Z > 56$ removed (presented in Sections~$\ref{sec:Spectral analysis results - Epoch 2} - \ref{sec:Spectral analysis results - Epoch 4}$).

\par
To determine how sensitive our models are to the presence of lanthanide material, we varied \Xlanth\ in our best-fitting \AngIII\ composition profile models, across epochs $2 - 7$. We find that our models can tolerate a variation in \Xlanth\ of a factor $\lesssim 2$. The optimal lanthanide mass fraction varies slightly from epoch to epoch, but typically our \tardis\ models are insensitive to variations in \Xlanth\ within the range $0.03 \lesssim$~\Xlanth~$\lesssim 0.10$. Here we again note that our analysis is subject to systematic uncertainties introduced from relying on incomplete atomic data for these lanthanide species. Although we are now in a regime where most of the lanthanide material is singly ionised (for which we have \textsc{dream} data), this data is expected to be quite incomplete. However, since the \textsc{dream} atomic line lists are obtained experimentally, they are likely to preferentially feature stronger lines, which we na\"ively expect to dominate the spectrum. How much of an effect the many weaker lines that we have not accounted for would have cannot be quantitatively determined. As a result, our lanthanide mass fraction constraints may be higher than we would infer if we had access to a complete line list for all lanthanide species. Despite such uncertainties, we argue that we have placed meaningful constraints on \Xlanth. \cite{Domoto2021} compare model lightcurves for \gfo, one generated from an incomplete atomic line list, and the other generated from theoretical line lists. They showed that the incompleteness of atomic data does not substantially affect their model lightcurves or synthetic spectra at +1.5 days. Additionally, \cite{Kasen2017} show that they can broadly reproduce the lightcurves and SEDs of \gfo\ with two components, one of which has a lanthanide mass fraction, \Xlanth~$= 10^{-1.5} \simeq 0.03$, which is consistent with the value we invoke for our modelling ($0.03 \lesssim$~\Xlanth~$\lesssim 0.10$). \cite{Coughlin2018} also present two-component models which they use to model both the lightcurve and early spectra of \gfo. From their models, they infer $\log_{10}$(\Xlanth)~$= -1.61^{+0.96}_{-1.04}$ (\Xlanth~$\simeq 0.025$) from their lightcurve analysis, and $\log_{10}$(\Xlanth)~$= -1.52^{+0.97}_{-0.98}$ (\Xlanth~$\simeq 0.030$) from their spectral modelling. Both of these values are similar to those invoked by \cite{Kasen2017}, and lie close to our inferred values.

\par
Considering all of the above, we reach the strong conclusion that without the heavy lanthanide elements present in some moderate quantity (\Xlanth\ $\simeq 0.05^{+0.05}_{-0.02}$), the \AngIII\ model spectra cannot produce enough absorption at wavelengths $\lesssim 7500$\,\AA\ to match the observations, from $+2.4 - 7.4$ days.

\par
The flux-calibrated \xsh\ spectra between $+2.4 - 4.4$\,d are distinctly red, with the bulk of the flux emitted at wavelengths \mbox{$\gtrsim 8000$\,\AA} \citep[as originally noted by][]{Chornock2017, Kasliwal2017, Pian2017}. Our analysis suggests that this is a direct result of the strong line absorption of the lanthanides. In our models it is the result of cerium, neodymium, samarium and europium in particular, but without complete atomic data, it is hard to positively identify which species will dominate, and which we can rule out. It is the absorption deficit in the blue that is the primary cause of the `NIR excess' and the apparent `red kilonova' appearance from +2.4\,d onward. Previous theoretical works \citep[e.g.][]{Barnes2013, Kasen2013, Tanaka2013, Fontes2015, Kasen2017, Wollaeger2018} had predicted that such line absorption by the complex lanthanide ions would be likely in kilonovae, and that it would significantly alter the spectrum shape. However, we note that our lanthanide mass fraction requirement is relatively modest (\Xlanth~$\simeq 0.05^{+0.05}_{-0.02}$), and we can reject a very lanthanide-rich composition.

\par
Therefore, we have evidence supporting a disjointed composition, in which the first \xsh\ spectrum, at +1.4\,d, requires a different composition to all the others, and the $+2.4 - 7.4$\,d spectra are reasonably well reproduced with a consistent abundance profile. The ejecta in the first $1 - 2$ days after merger appear to be dominated by a light, lanthanide-free component, travelling at high velocities. This is followed by another, slower moving component, with a modest (but significant) lanthanide mass fraction that contributes significantly to the observed spectra $\gtrsim 2$ days. This could be indicative of stratified ejecta. At very early times, a significant portion of the inner ejecta material will be optically thick. As a result, any emergent spectra will be dominated by continuum contribution, with the only deviations arising from the small amount of optically thin material at the outer edges of the ejecta. However, with increasing time, the ejecta material will expand and cool, such that we can `see' deeper into the ejecta, where the heavier elements would reside. Therefore, we would expect to see some composition evolution as the ejecta material expands and cools with time, lending credence to the idea of stratified material.

\par
Alternatively, this could be evidence in favour of the two-component model for BNS mergers \citep[e.g. as presented by][]{Kasen2017}, with the blue component dominating at +1.4\,d, and the red component dominating beyond $\gtrsim 2$\,d. This temporal separation (between $1 - 2$ days) of the blue and red components is also consistent with the photometric analysis of the lightcurve by \cite{Cowperthwaite2017} and \cite{Villar2017} in particular. If the spectra are indeed produced by two components, our \tardis\ ejecta velocities can be seen to broadly agree with the velocities expected for the dynamical and disk wind components at these epochs. As previously discussed in Section~\ref{sec:Introduction}, for dynamical ejecta, we expect ejecta speeds between \mbox{$0.2 - 0.3\,c$}, and for disk wind ejecta, speeds on the order of $\sim 0.1\,c$. In Table~\ref{tab:TARDIS model parameters}, we present the $v_{\rm min}$ values of our best-fitting \tardis\ models for the first four epochs, and these broadly agree with the notion of the earliest epoch being dominated by a blue ejecta component (dynamical) and subsequent epochs being dominated by a red ejecta component (disk wind). This would be consistent with the \cite{Kasen2017} model of the `squeezed dynamical' blue component and a red disk wind formed when the remnant collapses promptly to a black hole.
 
\par
In this work, we find agreement with the primary result presented by \cite{Watson2019}, where they identify \SrII\ as the source of the strong absorption feature between $\sim 7000 - 10000$\,\AA\ in the early spectra of \gfo. As we include all available atomic data for the second \rpro\ peak, we also agree with \cite{Watson2019} that this absorption is not due to neutral Cs or Te, as originally proposed by \cite{Smartt2017}. However, our conclusions differ from the other \tardis\ modelling conclusions of \cite{Watson2019}. For the +1.4\,d \xsh\ spectrum, we find good agreement between observation and our \tardis\ models by invoking a composition dominated by first \rpro\ peak elements. Specifically, we find that Zr absorption is required to match the blue end of the spectrum ($\lambda \lesssim 5000$\,\AA) and we produce quite a satisfactory fit across all wavelengths. The \cite{Watson2019} \tardis\ model fails to reproduce the observed spectrum blueward of the \SrII\ feature, being under-luminous at $\sim 6000$\,\AA, and lacking absorption below $\sim 4500$\,\AA. The subsequent \tardis\ models for the $+2.4 - 4.4$\,d spectra presented by \cite{Watson2019} are incomplete matches to the data, as they fail to reproduce the emission of the \SrII\ P-Cygni line, and the inclusion of the heavier \rpro\ elements (in solar abundance ratios) produces too much blue absorption. 

\par
This is not a criticism of the \cite{Watson2019} \tardis\ work, since their primary aim was to show that the \SrII\ line formation in a \tardis\ model produced consistent results with the simpler P-Cygni line formation calculation, and that conclusion appears quite solid. Here we have performed a more systematic and thorough analysis, and have considered the entire wavelength range, with consistent compositions and \tardis\ densities and velocities. Thus, we consider that we have confirmed the primary result of \cite{Watson2019}, but have gone further by illustrating that either a layered or two-component model is required, with a lanthanide mass fraction, \mbox{\Xlanth~$\simeq 0.05^{+0.05}_{-0.02}$} required from +2.4\,d onward, and a lanthanide-free composition \mbox{(\Xlanth~$\lesssim 5 \times 10^{-3}$)} required at +1.4\,d and earlier.

\par
If the prominent $\sim 7000 - 10000$\,\AA\ feature is indeed produced by the \SrII\ triplet, then, as the ejecta evolve to become optically thin, we might expect the [\SrII] doublet to appear in emission (at 6738.4 and 6868.2\,\AA, in air). This is the equivalent of the strong [\CaII] 7291.5 and 7232.9\,\AA\ doublet (in air) that is prominent in the nebular spectra of many supernovae. The electronic configurations of these ions are similar (they are both Group 2 elements) and the forbidden doublet is produced from the de-excitation of the lower levels of the \SrII\ triplet (4p$^6$4d~$^2$D to ground). This is not observed in the \xsh\ spectra, and we discuss this problem in more detail in \paperII.

\subsection{Ionisation and atomic data uncertainties} \label{sec:Discussion and interpretation - Ionisation and atomic data uncertainties}
\par
Here we explore how ionisation uncertainties may affect our model parameters. We generally have judged the quality of our fits based on the required line-blanketing at wavelengths $\lesssim 6000$\,\AA, and the fit to the \SrII\ absorption feature. At +1.4\,d, with our best-fitting temperature and density, it is important to note that almost all of the Sr is doubly ionised (Sr\III). The two models which most closely match the data at +1.4\,d (\AngII\ and $1^{\rm st}$ $r$-peak) contain \mbox{$4.7 \times 10^{-5}$ and $6.0 \times 10^{-5}$\,\msun} of Sr in the line-forming region, only $\sim 0.3$ and 0.2 per cent of which is \SrII\ (see Table~\ref{tab:TARDIS model parameters}). However, these two best-fitting models for the first epoch contain much more total Sr mass than the subsequent models, due to the relatively high temperature in the line-forming region ($T = 4500$\,K), and the low fraction of Sr in the singly ionised state. The relative mass fraction of Sr required in our best-fitting composition profiles drops \mbox{from $\sim 20$ and 75} per cent to $\sim 3$ per cent, between the first epoch and the next three (see Table~\ref{tab:TARDIS model parameters}). Therefore, we should be careful with inferring the details of the composition in the first epoch primarily from the appearance of \SrII\ absorption, when \SrII\ is a trace ion in the ejecta. 

\par
The exact amount of \SrII\ in the line-forming region is strongly sensitive to small variations in temperature (and to a lesser extent, density). Reducing $T$ or increasing $\rho_0$ leads to more \SrII\ present in the model (and less Sr\III, since both of these modifications act to reduce the ionisation of the ejecta material). Uncertainty in either of these model parameters, or in the ionisation approximation in our \tardis\ models, strongly affect the relative ratio of \SrII/Sr. Without more rigorous constraints on the ionisation state of the plasma, we cannot determine whether our ionisation approximation within \tardis\ is sufficiently accurate. For these models, we suffer from the lack of additional features that we can tie to other species. If we could constrain the identification of one (or both) of the strong NIR emission features to some species, then we may be able to better infer properties of the plasma from how highly ionised the ejecta needs to be to produce this species.

\par
\begin{figure}
    \centering
    \includegraphics[width=\linewidth]{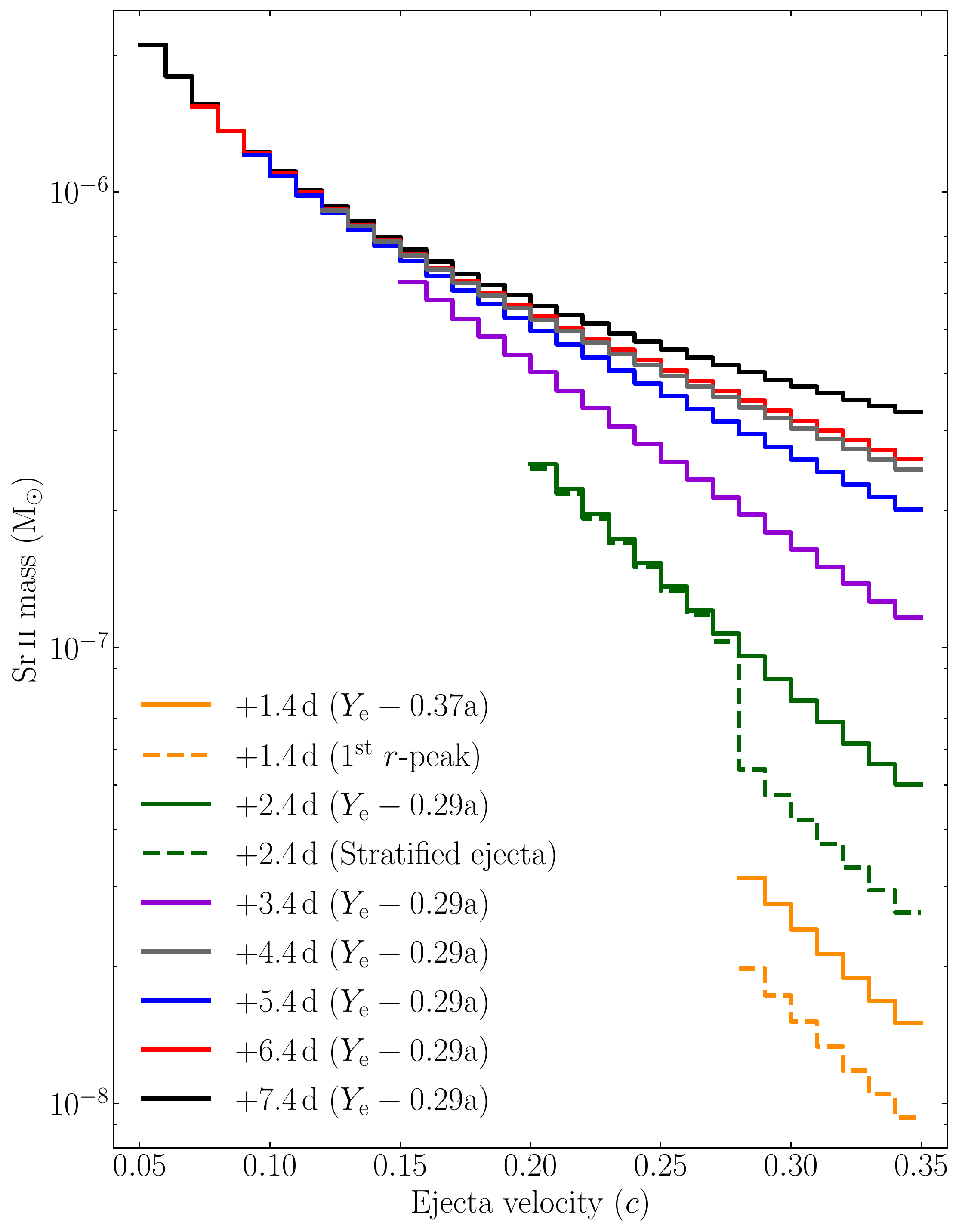}
    \caption{
        Distribution of \SrII\ mass in all of the \tardis\ models presented in this work. The models have been binned into shells with width $0.01\,c$, and so the masses represent the total \SrII\ mass in these shells. The total masses of \SrII\ for the $+1.4 - 4.4$\,d models are presented in Table~\ref{tab:TARDIS model parameters} (apart from the +2.4\,d stratified ejecta model, which has \mbox{\MSrII\ = $1.6 \times 10^{-6}$\,\msun}; see Section~\ref{sec:Discussion and interpretation - Stratified ejecta}). The total \SrII\ masses for the +5.4, +6.4 and +7.4\,d models are $1.4 \times 10^{-5}$, $1.8 \times 10^{-5}$ and $2.3 \times 10^{-5}$\,\msun, respectively.
    }
    \label{fig:SrII masses}
\end{figure}

This (potential) uncertainty in the ionisation of Sr is not as significant an issue for the later models, where the mass of \SrII\ is comparable to, or larger than, the mass of Sr\III. Despite this, we are reasonably confident in the estimate of the amount of \SrII\ required to produce the observed absorption feature between $\sim 7000 - 10000$\,\AA\ in the +1.4\,d spectrum, but with some uncertainty in the total Sr mass. It is clear that the mass of \SrII\ increases with time across epochs~$1 - 7$ (see Table~\ref{tab:TARDIS model parameters} and Figure~\ref{fig:SrII masses}). In Figure~\ref{fig:SrII masses}, we plot the \SrII\ mass distributions for all of our best-fitting \tardis\ models presented in this work. The \SrII\ mass has been binned into shells, with width $0.01\,c$, and so the plot illustrates the total mass of \SrII\ in each shell in the model. The mass of \SrII\ increases with time, both in each shell (due to the drop in ionisation, which leads to a drop in the amount of Sr\III\ in the model), and across the entire line-forming region (primarily due to the recession of the inner boundary).

\par
Here we note the one exception to this trend. In our fifth epoch model, the ionisation of Sr is higher than our model for the fourth epoch, which leads to less \SrII\ material in the outer velocity bins. This is simply a result of our model temperature remaining constant between epochs four and five ($T = 3200$\,K). Since the ionisation of the ejecta material is sensitive to changes in $T$ and $\rho_0$, and our model temperature remains constant, the model ionisation is impacted by the small decrease in density, which acts to slightly increase the ionisation. Considering all of the above, this indicates that, despite our uncertainty in the ionisation of Sr (especially at the first epoch), our \SrII\ mass evolution with time behaves as expected. We also conclude that, since \SrII\ is a trace element in the +1.4\,d model, the uncertainty in the ionisation state of the element means that we cannot be definitive in ruling out a solar abundance ratio of Sr~:~Y~:~Zr (as discussed in Section~\ref{sec:Spectral analysis results - Epoch 1 - First r-process peak model}). A better understanding of the ionisation of Sr, beyond a simple LTE approximation, would increase the reliability of our Sr mass estimate (and therefore our total mass estimate) at early times.

\par
The \textsc{dream} database is an extremely useful source of experimentally verified lanthanide data. Although it may be incomplete, the important contribution from the lanthanides is expected to be at blue wavelengths ($\lesssim 6000$\,\AA), due to line blanketing from a multitude of similar strength lines between low-lying levels \citep{Barnes2013,Kasen2013,Tanaka2013,Tanaka2020}. Similar behaviour is expected for all lanthanide species, and so, having incomplete data for some of these species (at UV and optical wavelengths) may not cause a significant problem in our modelling. It is a multitude of lines that create the depressed continuum, rather than a small number of transitions shaping broad line profiles. Therefore, we expect our constraint on the total mass fraction of lanthanide material to be reasonably secure, but do not place any particular focus on the mass fractions of individual species.

\par
The largest source of uncertainty in our analysis is the lack of complete and reliable atomic data. There is a serious lack of any data for some potentially important elements in our modelling. Specifically, we are lacking good atomic data for the second \rpro\ peak elements that are dominant in our \AngIII\ composition profile, e.g. Ru, Sn and Te. We also suffer from incomplete data for many others. For example, only three of the lanthanide species that we included from \textsc{dream} have data for transitions with $\lambda > 1$\,\micron\ (La\I, Pr\IV\ and Yb\II). This prevents us from exploring any NIR modelling with the majority of the lanthanide species. Further spectroscopic modelling of \gfo\ (and future KNe) would benefit from new atomic data calculations, with a particular focus on extending current data sets beyond 1\,\micron, and on calibration, so the data have precise wavelengths and transition strengths. Without complete and well-calibrated atomic data for all of the species of interest in BNS mergers, it is impossible to make a definitive line identification for observed features within the spectra of KNe. Additionally, accurate line identification is affected by the large expansion velocities associated with the ejecta of KNe. These high velocities make line-blending in the observed spectra common, making definitive line identification difficult, which will in turn affect accurate line identification studies.

\subsection{Stratified ejecta} \label{sec:Discussion and interpretation - Stratified ejecta}
\par
For all \tardis\ models, we can obtain good agreement with the data using a continuous density profile (where \mbox{$\rho_0 = 4 \times 10^{-15}$\gcm}; see Table~\ref{tab:TARDIS model parameters}) and a uniform composition for each individual epoch of interest, albeit with a different requirement for the composition for the first epoch. We opted against over-complicating our \tardis\ models, and so did not pursue further investigation of multi-zone and stratified models. We took this approach as there are already a large number of free parameters in our model ejecta properties (including composition). Attempting to further fine-tune our composition at different velocities, while obtaining meaningful and objective conclusions would be unrealistic, given the nature of our analytic approach. We acknowledge that our approach may seem formally inconsistent, where we invoke uniform compositions in our models, but with a different uniform composition for the first epoch. Because of this, here we will demonstrate that it is possible to obtain an equivalently `good' fit using a two-zone, stratified ejecta composition approach.

\par
\begin{figure}
    \centering
    \includegraphics[width=\linewidth]{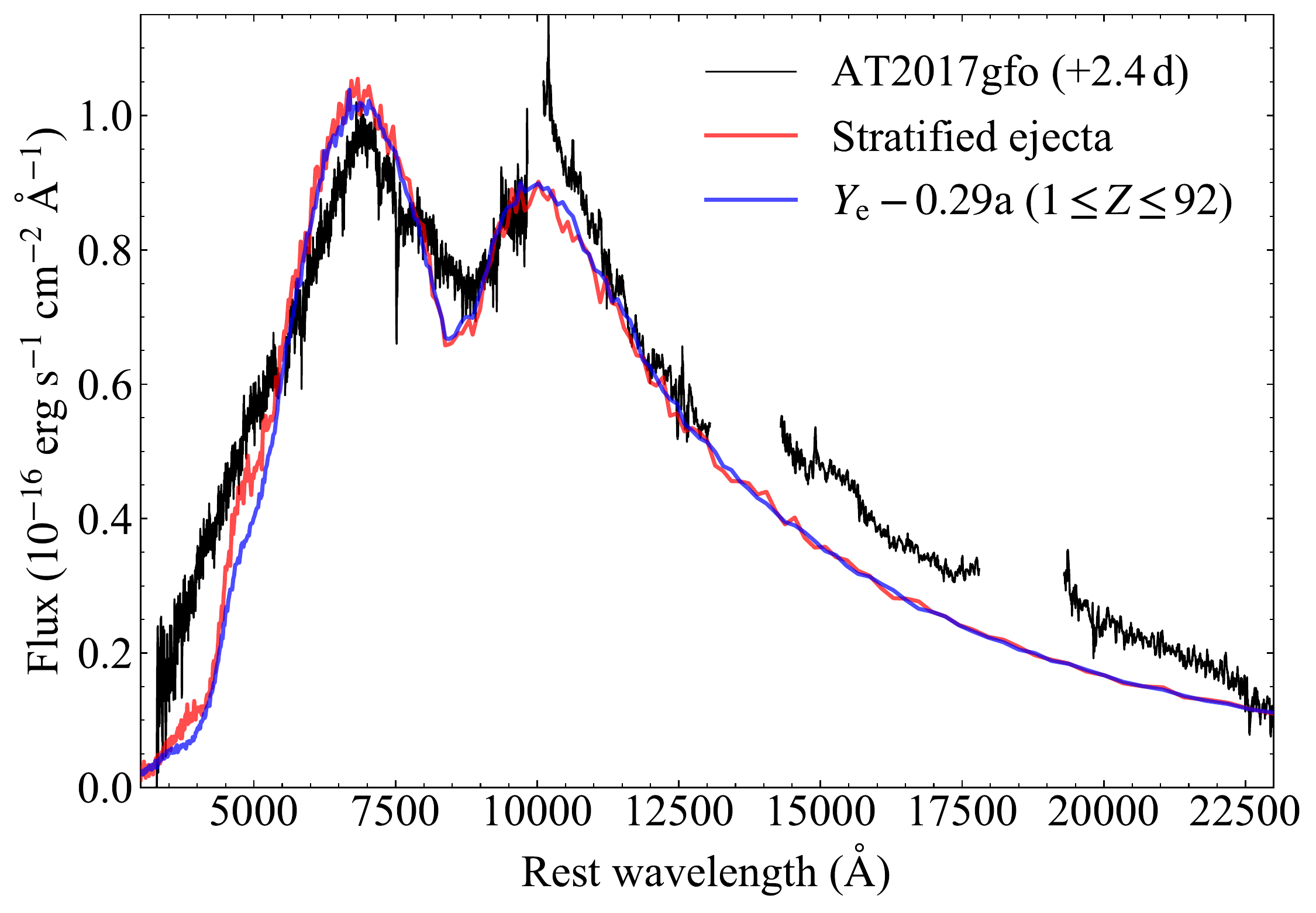}
    \caption{
        +2.4\,d \xsh\ spectrum of \gfo, plotted with our best-fitting \AngIII\ model (blue), and a model with stratified density and composition (red).
    }
    \label{fig:2.4d Stratified model}
\end{figure}

In Figure~\ref{fig:2.4d Stratified model}, we present one such example of a stratified ejecta model, for the +2.4\,d spectrum. Here we have used the \AngIII\ composition for ejecta in the velocity range \mbox{$0.20 \leq v < 0.28\,c$}. This velocity range corresponds to the recession of the inner boundary velocity between our best-fitting models for the first and second epochs, presented in Sections~\ref{sec:Spectral analysis results - Epoch 1} and \ref{sec:Spectral analysis results - Epoch 2}. We then used our \mbox{$1^{\rm st}$ $r$-peak} model composition for the ejecta in the higher velocity range, \mbox{$0.28 \leq v \leq 0.35\,c$}. For velocities $\geq 0.28\,c$, this model has an identical composition to our best-fitting model for the +1.4\,d \xsh\ spectrum of \gfo, and for velocities $< 0.28\,c$, it has an identical composition to our best-fitting model for the +2.4\,d \gfo\ spectrum (and all subsequent spectra). Also plotted in Figure~\ref{fig:2.4d Stratified model} is our best-fitting \AngIII\ model presented in Section~\ref{sec:Spectral analysis results - Epoch 2}, for comparison. Clearly, both the uniform single zone model, and this stratified model, are equally good at replicating the observations at this epoch.

\par
However, to get the stratified model to agree with the observed spectrum, we needed to reduce the density of the ejecta with $v \geq 0.28\,c$ by a factor of 10. We require a value of \mbox{$\rho_0 = 4.0 \times 10^{-16}$\gcm}, which is a factor of 10 lower than the value of $\rho_0$ used for the equivalent line-forming region for the first epoch, $1^{\rm st}$ $r$-peak model. This reduction in $\rho_0$ is necessary, as our outer composition contains a large quantity of Sr, and, due to the cooler ejecta temperature at this second epoch (3600\,K, versus 4500\,K for the first epoch model), it is now no longer completely dominated by Sr\III. The line-forming region in this stratified model contains much more \SrII\ than is required to reproduce the \mbox{$\sim 7000 - 10000$\,\AA} absorption feature. Hence, we found it necessary to drop the overall density to obtain a satisfactory match.

\par
This model indicates that it would be possible to replicate the early observational data using this approach. However, the disjoint in both the composition and the density profile that we have invoked to achieve consistency across the first and second epochs, makes our model choices somewhat arbitrary. The fact that a stratified model with a discontinuous density profile can also reproduce the observed spectra indicates that the physical properties of the line-forming region (including its composition) are changing with time. However, we cannot distinguish whether this is just ejecta with non-uniform composition (i.e. stratified material), or physically distinct ejecta components (i.e. the two-component KN model).

\par
Finally, it is worth highlighting that, although we originally allowed $\rho_0$ to vary as a free parameter to improve fit quality at each epoch, we found that our best-fitting models all have a consistent value ($\rho_0 = 4 \times 10^{-15}$\gcm). The fact we converged towards a consistent value across all epochs increases the credibility of our modelling efforts.

\section{Conclusions} \label{sec:Conclusions}
\par
The main aim of this work was to generate a sequence of spectroscopic models, using \tardis, that could accurately reproduce the evolution of the early spectra of the kilonova, \gfo. We compared these \tardis\ models to precisely calibrated $0.3 - 2.5$\,\micron\ high-quality \xsh\ spectra, taken on a daily basis over the first ten days of the evolution of this exceptional transient. To constrain the composition of the ejecta, and attempt identification of individual elements and ions, we used compositions reflective of realistic nucleosynthetic trajectories calculated for a binary neutron star merger.

\par
No specific information on the ejecta composition could be extracted at very early times ($\lesssim 1$ day) due to the lack of absorption features at the wavelengths covered by the spectra. However, we predict strong absorption in the UV, and so we determine that UV observations of future events may be able to provide useful information on the composition of the first material ejected.

\par
The first \xsh\ spectrum, taken +1.4\,d after merger, is significantly bluer than all subsequent epochs. We find good agreement to the data with a \tardis\ model with ejecta temperature, $T = 4500$\,K, and a composition that is dominated by the elements at the first \rpro\ peak. We require an inner boundary velocity, $v_{\rm min} = 0.28\,c$, at the base of the line-forming region. We reproduce and confirm the main result of \cite{Watson2019}, namely that the broad absorption feature between $\sim 7000 - 10000$\,\AA\ can be quantitatively explained by the \SrII\ NIR triplet. We find that it is impossible to reproduce the shape of the spectrum taken at +1.4\,d with lanthanide material present in any significant quantity \mbox{(lanthanide mass fraction, \Xlanth\ $\lesssim 5 \times 10^{-3}$)}. The multitude of transitions from low-lying levels of these species result in strong absorption at blue wavelengths ($\lambda \lesssim 6000$\,\AA), precluding their presence in the line-forming region at this early epoch. Formally, the composition which produces the closest agreement with the data is one composed entirely of first \rpro\ peak elements (75 per cent Sr and 25 per cent Zr). While this deviates somewhat from the solar ratio of these elements, the uncertainty in the ionisation states (particularly \SrII), and the uncertainty in the available atomic data, prevent stronger conclusions. We propose that a composition dominated by first \rpro\ peak elements broadly explains this first \xsh\ spectrum.

\par
However, the second and all subsequent \xsh\ spectra require quite a different ejecta composition. The SED changes significantly between 1.4 and 2.4 days, becoming rapidly redder and peaking at $\sim 1$\,\micron. We require a modest amount of lanthanide material \mbox{(\Xlanth\ $\simeq 0.05^{+0.05}_{-0.02}$)} to reproduce the observed flux deficit at wavelengths $\lesssim 7000$\,\AA\ with line absorption. We interpret the prominent feature between $\sim 7000 - 10000$\,\AA\ as a P-Cygni line from the \SrII\ triplet. We can reproduce this strong line profile in \tardis, and simultaneously fit the blue deficit with a realistic composition profile extracted from a hydrodynamical simulation of a BNS merger, with an intermediate electron fraction that produces a mix of first \rpro\ peak species, and heavier material (our \AngIII\ composition). 

\par
The spectra at +3.4 and +4.4\,d can be replicated well with exactly the same \AngIII\ composition, indicating that there is no strong deviation in composition from $\sim 2 - 5$ days post-merger. A single zone, uniform composition is sufficient to satisfactorily reproduce the spectral evolution, while maintaining a consistent density profile, and a receding photosphere and cooler ejecta with increasing time (as expected for homologous expansion). We can reasonably reproduce the +5.4 and 6.4\,d spectra by consistent forward modelling, but we note that by +7.4\,d, the transient appears to no longer be in a photospheric regime, and so \tardis\ is no longer suited to model the spectra.

\par
We find a strong quantitative argument for a disjoint in the composition of the ejecta. This may suggest either stratified material, or observations of two distinct components of ejecta material. We find that a high velocity `blue kilonova' component is prominent only for the first $\lesssim 1 - 2$ days, and has a composition dominated by first \rpro\ peak material (Sr and Zr), with virtually no lanthanide species \mbox{(\Xlanth\ $\lesssim 5 \times 10^{-3}$)}. Additionally, we find that all subsequent spectra show strong line blanketing below $\lesssim 7500$\,\AA, which is due to the lanthanides being present with a total mass fraction, \mbox{\Xlanth\ $\simeq 0.05^{+0.05}_{-0.02}$}. This represents a `red kilonova' component, with signatures of line blanketing by cerium (Ce\II), neodymium (Nd\II\ and Nd\III), samarium (Sm\II) and europium (Eu\II), among others. Thus, the red kilonova is distinguished more by a blue flux deficit, rather than a near-infrared excess.

\par
The largest uncertainty in our analysis arises from the lack of a complete atomic data set. We again stress that without complete and well-calibrated atomic data for all of the species of interest in BNS mergers, it is impossible to make a definitive line identification for observed features within the spectra of KNe. For the compositions we favour for the ejecta of \gfo\ in this work, the elements we most pertinently need better atomic data are those belonging to the second \rpro\ peak, e.g. ruthenium, tin and tellurium.

\section*{Acknowledgements}
We thank the anonymous referee for the useful feedback and helpful comments.
We thank Michael McCann for providing access to his platinum atomic data.
SAS and SJS acknowledge funding from STFC Grants ST/P000312/1 and ST/T000198/1.
AB acknowledges support by the European Research Council (ERC) under the European Union’s Horizon 2020 research and innovation programme under grant agreement No.\ 759253, and support by Deutsche Forschungsgemeinschaft (DFG, German Research Foundation) - Project-ID 279384907 - SFB 1245 and DFG - Project-ID 138713538 - SFB 881 (“The Milky Way System”, subproject A10) and support by the State of Hesse within the Cluster Project ELEMENTS.
SG acknowledges financial support from F.R.S.-FNRS (Belgium). This work has been supported by the Fonds de la Recherche Scientifique (FNRS, Belgium) and the Research Foundation Flanders (FWO, Belgium) under the EOS Project nr O022818F.
CHIANTI is a collaborative project involving George Mason University, the University of Michigan (USA), University of Cambridge (UK) and NASA Goddard Space Flight Center (USA).
This research made use of \tardis, a community-developed software package for spectral synthesis in supernovae. The development of \tardis\ received support from the Google Summer of Code initiative and from ESA's Summer of Code in Space program. \tardis\ makes extensive use of Astropy and PyNE.
We are grateful for use of the computing resources from the Northern Ireland High Performance Computing (NI-HPC) service funded by EPSRC (EP/T022175).
Based on observations collected at the European Southern Observatory (ESO) under programmes
099.D-0376, 
099.D-0382, 
099.D-0622 
and 099.D-0191, 
and made available through the ESO Science Archive Facility (\url{http://archive.eso.org}).
We made use of the flux-calibrated versions of the \xsh\ spectra publicly available through ENGRAVE.

\section*{Data Availability}
All \tardis\ model spectra that have been presented, as well as an extended version of Table~\ref{tab:Ye bins (top 10)} that contains the complete abundances for the composition profiles considered in this work are available, and can be accessed from the Queen's University Belfast \href{https://pure.qub.ac.uk/en/datasets/dataset-for-the-paper-modelling-the-spectra-of-the-kilonova-at201}{Pure research portal}.

\bibliographystyle{mnras}
\bibliography{ref} 


\bsp	
\label{lastpage}
\end{document}